\documentclass[lettersize,journal]{IEEEtran}
\usepackage{amsmath,amsfonts}
\usepackage{algorithmic}
\usepackage{algorithm}
\usepackage{array}
\usepackage{subfigure}
\usepackage{hyperref} 
\usepackage{textcomp}
\usepackage{stfloats}
\usepackage{url}
\usepackage{verbatim}
\usepackage{makecell}
\usepackage{multirow}
\usepackage{graphicx}
\usepackage{cite}
\usepackage{bbding}
\usepackage{amssymb}
\usepackage{color}
\usepackage{hyperref}
\usepackage{xcolor}

\begin{document}

\title{Multi-level Temporal-channel Speaker Retrieval for Zero-shot Voice Conversion}

\author{Zhichao~Wang,
    Liumeng~Xue,
    Qiuqiang~Kong,
    Lei~Xie,~\IEEEmembership{Senior Member,~IEEE,}
    \\Yuanzhe Chen,
    Qiao Tian,
    Yuping Wang
  % <-this % stops a space
%  \thanks{Corresponding author: Lei Xie}
  
% \thanks{Zhichao Wang, Liumeng Xue, and Lei Xie are with the School of Computer Science, Northwestern Polytechnical University, Xi’an 710072, China. Email: zcwang\_aslp@mail.nwpu.edu.cn (Zhichao Wang), lmxue@nwpu-aslp.org (Liumeng Xue), lxie@nwpu.edu.cn (Lei Xie)}

% \thanks{Qiuqiang Kong, Yuanzhe Chen, Qiao Tian, and Yuping Wang are with the Speech, Audio, and Music Intelligence (SAMI) Group, ByteDance, Shanghai 200233, China. Email: 
% kongqiuqiang@bytedance.com (Qiuqiang Kong), chenyuanzhe@bytedance.com (Yuanzhe Chen), tianqiao.wave@bytedance.com (Qiao Tian), wangyuping@bytedance.com (Yuping Wang)}
    
    % 13-Dec-202312-May-2024
 \thanks{Corresponding author: Lei Xie}
  
\thanks{Zhichao Wang, Liumeng Xue, and Lei Xie are with the School of Computer Science, Northwestern Polytechnical University, Xi’an 710072, China. Email: zcwang\_aslp@mail.nwpu.edu.cn (Zhichao Wang), lmxue@nwpu-aslp.org (Liumeng Xue), lxie@nwpu.edu.cn (Lei Xie)}

\thanks{Qiuqiang Kong, Yuanzhe Chen, Qiao Tian, and Yuping Wang are with the Speech, Audio, and Music Intelligence (SAMI) Group, ByteDance, Shanghai 200233, China. Email: 
kongqiuqiang@bytedance.com (Qiuqiang Kong), chenyuanzhe@bytedance.com (Yuanzhe Chen), tianqiao.wave@bytedance.com (Qiao Tian), wangyuping@bytedance.com (Yuping Wang)}

    }
\markboth{Journal of \LaTeX\ Class Files,~Vol.~14, No.~8, August~2021}%
{Shell \MakeLowercase{\textit{et al.}}: A Sample Article Using IEEEtran.cls for IEEE Journals}


\maketitle

\begin{abstract}
 % Zero-shot voice conversion (VC) aims to convert speech into the voice of any desired speaker, using just one recording of the speaker, without requiring additional model updates. Typical methods utilize speaker representation extracted from a pre-trained speaker verification (SV) model or jointly learned during training to achieve zero-shot VC. However, the process of speaker modeling involves redundant information in the temporal and channel dimensions of the speech. It hampers the models from accurately representing unseen speakers whose speech does not appear in the training dataset. 

 Zero-shot voice conversion (VC) converts source speech into the voice of any desired speaker using only one utterance of the speaker without requiring additional model updates. Typical methods use a speaker representation from a pre-trained speaker verification (SV) model or learn speaker representation during VC training to achieve zero-shot VC. However, existing speaker modeling methods overlook the variation of speaker information richness in temporal and frequency channel dimensions of speech. This insufficient speaker modeling hampers the ability of the VC model to accurately represent unseen speakers who are not in the training dataset. In this study, we present a robust zero-shot VC model with \textit{\textbf{m}ulti-level \textbf{t}emporal-\textbf{c}hannel \textbf{r}etrieval}, referred to as MTCR-VC. Specifically, to flexibly adapt to the dynamic-variant speaker characteristic in the temporal and channel axis of the speech, we propose a novel fine-grained speaker modeling method, called \textit{\textbf{t}emporal-\textbf{c}hannel \textbf{r}etrieval (TCR)}, to find out \textit{when} and \textit{where} speaker information appears in speech. It retrieves variable-length speaker representation from both temporal and channel dimensions under the guidance of a pre-trained SV model. Besides, inspired by the hierarchical process of human speech production, the MTCR speaker module stacks several TCR blocks to extract speaker representations from multi-granularity levels. Furthermore, we introduce a cycle-based training strategy to simulate zero-shot inference recurrently to achieve better speech disentanglement and reconstruction. To drive this process, we adopt perceptual constraints on three aspects: content, style, and speaker. Experiments demonstrate that MTCR-VC is superior to the previous zero-shot VC methods in modeling speaker timbre while maintaining good speech naturalness.

% is superior to the previous VC methods

% achieves superior performance in modeling speaker timbre while maintaining good speech naturalness.

 % In this study, we propose a robust zero-shot VC model named MTCR-VC, which is designed to achieve high speaker similarity for unseen speakers. Specifically, considering the dynamic-variant speaker characteristic in the temporal and channel axis of the speech, taking advantage of the speaker discriminative knowledge learned in the SV, \textit{\textbf{t}emporal-\textbf{c}hannel \textbf{r}etrieval (TCR)} is proposed to retrieve speaker timbre from target speaker's speech under the guidance of SV's speaker embedding by temporal-channel attention mechanism. With the hierarchical nature of speech production which results in changes of speaker timbre information in different granularities, the MTCR speaker module which stacks several TCR blocks is utilized to extract speaker representations from multi-granularity levels and fuse them to corresponding decoder layers like U-net.  This method allows the model to flexibly adjust attention on temporal and channel dimensions according to the richness of speaker information. Furthermore, to drive the process of speech reconstruction and disentanglement, a cycle-based training strategy with perceptual constraints is introduced to simulate the zero-shot VC process during training. Experiments demonstrate that MTCR-VC achieves new state-of-the-art performance for zero-shot VC in speaker similarity.
\end{abstract}


\section{Introduction}
\label{sec:intro}
Voice conversion (VC) aims to convert the source speech to a target speaker without changing the linguistic content. VC has been deployed to many applications, including dubbing, live broadcast, voice anonymization, and pronunciation correction. 
In recent years, neural networks-based VC systems, such as generative adversarial network (GAN)~\cite{GANHsu2017VoiceCF}, variational auto-encoder (VAE)~\cite{VAEHsu2016Voicevae}, and recognition-synthesis autoencoder framework~\cite{PPGSun2016PhoneticPF} are widely adopted. However, these VC models can only convert source speech to that of predefined target speakers. Meanwhile, these VC models usually require many speech recordings of the target speaker for model training. As data collection is expensive and time-consuming, building a high-quality VC system with minimal data requirements is more practical for real-world applications. Consequently, \textit{zero-shot VC}, also called \textit{any-to-any VC}, which aims to convert source speech to that of any speaker given only one utterance of the speaker, has drawn much attention recently. In this paper, we focus on the problem of zero-shot VC.
% In general, according to the speakers involved~\cite{liu2021any}, VC approaches can be categorized into one-to-one, many-to-one, many-to-many, any-to-many and any-to-any. Specifically, one-to-one VC represents that the VC model is capable of converting speech between two specific speakers. Conventional VC approaches focus on one-to-one VC, which aligns parallel data between the source and target speakers and learns a frame-wise mapping~\cite{GMMToda2007,NNSun2015}. Recent studies mainly use non-parallel data for more practical many-to-many voice conversion, where generative adversarial network (GAN)~\cite{maskcyclegan,stargan} and variational auto-encoder (VAE)~\cite{GANHsu2017VoiceCF} are widely adopted. With the help of text supervision, the recognition-synthesis autoencoder framework is able to achieve any-to-many VC, in which phonetic posteriorgram (PPG) or neural bottleneck feature (BNF)~\cite{PPGSun2016PhoneticPF,wang21g_interspeech}, computed from an automatic speech recognition (ASR) model, is adopted as a speaker-independent intermediate representation. 
% But these VC models can only convert source speech to that of predefined target speakers. Meanwhile, these VC models usually require a large number of speech recordings of the target speaker for model training. As data collection is expensive and time-consuming, building a high-quality VC system with minimal data requirements is more practical for real-world applications. 

As only one utterance is available, one of the key challenges in zero-shot VC is effectively capturing the target speaker's timbre to achieve good speaker similarity in the converted speech. Using a look-up table (LuT) to represent speaker identity~\cite{maskcyclegan,stargan,wang21g_interspeech} is a common practice but is limited to predefined speakers in the training data. Many recent studies attempt to directly capture speaker timbre from the limited speech of the target speaker. An intuitive approach is to leverage a pre-trained robust speaker verification (SV) model to extract utterance-level speaker representation~\cite{autovcqian2019autovc,speechsplit,speakeraware,mediumvc}. However, the SV model trained on many speakers and recording conditions is optimized for speaker classification but not for the \textit{perceptual} speaker similarity in the sense of human hearing. Instead of using an external SV model, many studies~\cite{INchou2019oneshot, AgainVC, SIGVC, avqvc, contrastive,nansy,VQMIVC,SRDVC,MAP,CAVC,du21_interspeech} decompose speech into linguistic content and speaker timbre, and even speaking style, aiming to separate utterance-level speaker timbre from other speech components. 
% For instance, instance normalization~\cite{INchou2019oneshot,AgainVC}, mutual information (MI)~\cite{VQMIVC,SRDVC}, adversarial training~\cite{SIGVC}, and information perturbation~\cite{speechsplit,nansy} are commonly used to wipe out the correlation between speaker timbre and other speech components. 
In the above approaches, speaker timbre is considered static and time-independent and modeled as a single coarse-grained fixed-length vector. To capture the dynamically varying speaker characteristics within an utterance, some recent studies~\cite{VQVC+,FragmentvcAVLin2021,retriever,attbasedzsl,Xuli_interspeech} focus on modeling more fine-grained speaker representation, which extracts speaker timbre from multiple aspects, covering multi-level and time-varying speaker representation. Based on the U-net~\cite{unet} structure, Li et al.~\cite{unetts}, Wu et al.~\cite{VQVC+}, and Li et al.~\cite{Xuli_interspeech} extract utterance-level speaker representations from multiple stacking layers and feed them to corresponding decoder layers. To access time-varying speaker information, some studies~\cite{attbasedzsl,FragmentvcAVLin2021,retriever} extract variable-length speaker representation and fuse it into the converted speech according to the content-based alignment between source speech and target speaker speech.

While the above progress has been made, existing methods of modeling fine-grained target speaker timbre are insufficient to capture the dynamic variation characteristic of speaker timbre.
The above speaker representation is usually extracted from the time-frequency space of speech, where a speech spectrogram includes both \textit{temporal} and frequency \textit{channel} dimensions. There are differences in speaker timbre richness in both temporal and channel dimensions of speech. Insufficient speaker modeling in these two dimensions may cause the unstable performance of the
VC model to represent unseen speakers. Moreover, speech production research~\cite{audio} shows that the frequency distributions of speech from different speakers lead to varying speaker timbre information in frequency channels~\cite{liu2022mfa, multi-frequency, ECAPA_TDNN}. Speech content, such as vowels, consonants, and para-linguistic features, carry distinct speaker timbre information reflected in temporal and frequency channel dimensions while silent speech segments apparently convey no speaker timbre information~\cite{attbasedzsl}. On the other hand, the human speech production mechanism is \textit{hierarchical}~\cite{human_voice, audio} in nature, from long-term airflow generation to fine-grained phoneme-related articulator movements and vocal filtering. 
% is produced by three main steps hierarchically~\cite{human_voice,audio}, from long-term airflow generation to fine-grained phoneme-related changes happening in the lungs, the vocal folds, and the articulators, 
Speaker-related information thus varies at different stages of speech production with different temporal and channel granularities. Finally, different speech factors such as linguistic content, speaking style, and speaker timbre are highly entangled in speech, and
factor disentanglement is necessary for better speaker timbre extraction. Compared with utterance-level speaker modeling, time-varying fine-grained speaker modeling can transfer more speaker information from the target speaker's speech to the converted speech. But this exacerbates the difficulty of disentanglement~\cite{attbasedzsl,FragmentvcAVLin2021}.

To address the aforementioned problems in zero-shot VC, we propose a novel fine-grained speaker modeling method called \textit{\textbf{t}emporal-\textbf{c}hannel \textbf{r}etrieval~(TCR)}, which captures the dynamic variation of speaker timbre in both temporal and channel dimensions. Specifically, with the help of the attention mechanism, the speaker embedding from the pre-trained SV model is used as a \textit{query} to \textit{retrieve} speaker timbre information in variable-length speaker representation from the target speaker. Inspired by the hierarchical nature of speech production, we employ \textit{\textbf{m}ulti-level TCR} (MTCR) in an encoder-decoder based U-net structure for voice conversion~\cite{unet} where fine-grained speaker information in different granularities is retrieved from different encoder layers. In the decoder, these multi-level speaker representations are then fused with content and style representations from the source speech to generate target speech. Furthermore, to perform better speech disentanglement, we introduce a \textit{cycle-based} training strategy to simulate zero-shot inference in a recurrent fashion. perceptual constraints~\cite{wang2022delivering} on three aspects, including content, style, and speaker, are adopted to drive this process. 
Zero-shot voice conversion experiments show that the proposed MTCR-VC approach generalizes well to cross-set speakers with superior speaker similarity. Audio samples can be found on our demo page.\footnote{\url{https://kerwinchao.github.io/demo_zslvc/}}

The main contributions of this work are as follows:
\begin{itemize}
    \item We propose a novel multi-level framework, MTCR-VC, for fine-grained speaker modeling in zero-shot VC. The \textit{multi-level temporal-channel retrieval} is designed to extract fine-grained speaker representations under the query of SV's speaker embedding in both temporal and channel dimensions and hierarchically capture speaker timbre from different granularities.
    \item We design a cycle-based training strategy to perform better speech disentanglement and reconstruction in the zero-shot scenario by simulating the zero-shot process with perceptual constraints.
\end{itemize}

The rest of this paper is organized as follows. Section~\ref{sc:related work} reviews the voice conversion and typical zero-shot VC framework, and then introduces the application of attention in VC. Section~\ref{sc:method} presents in detail the proposed MTCR approach for fine-grained speaker modeling. Section~\ref{sc:experiments} describes the experimental setup and Section~\ref{sc:results} presents the experimental results. Finally, Section~\ref{sc:conclusion} concludes the paper.

\section{Related work}
\label{sc:related work}

% This section reviews the typical zero-shot VC frameworks in the literature and then introduces our proposed MTCR-VC framework with a system overview.

This section will review related works on voice conversion and the typical zero-shot VC frameworks. We will also review the utilization of the attention mechanism in VC.

%introduce the commonly used zero-shot frameworks. Typical zero-shot VC framework can be categorized into two types based on the speaker modeling approach: utterance-level speaker modeling and fine-grained speaker modeling. Besides, we will provide an overview of our proposed MTCR-VC framework, which employs fine-grained speaker modeling.

% 在这个章节中我们将首先介绍zeroshot中常用的框架。按照说话人的建模方法我们将框架分为两类，包括句级说话人建模的模型框架，细颗粒度说话人建模的模型框架，其中细颗粒度建模包含。并且我们将介绍MTCR-VC的主要框架。

\subsection{Voice Conversion} 
VC approaches can be categorized into one-to-one, many-to-many, or any-to-any VC based on the number of speakers that the VC model supports. Conventional VC approaches focus on one-to-one VC, which align acoustic features of parallel data between a pair of source-target speakers and perform frame-wise mapping using Gaussian mixture models~\cite{GMMStylianou1998,GMMToda2007} and neural networks~\cite{NNDesai2009,NNSun2015}. With the high cost of parallel data, recent studies mainly use non-parallel data for voice conversion. GAN~\cite{maskcyclegan,stargan} and VAE~\cite{GANHsu2017VoiceCF} are further proposed to support many-to-many VC. With the help of text supervision, phonetic posteriorgram (PPG) and neural bottleneck feature (BNF)~\cite{PPGSun2016PhoneticPF} computed from an automatic speech recognition (ASR) model are assumed to be speaker-independent content representations. And thus PPG- and BN-based recognition-synthesis framework is widely used in any-to-many VC. Besides, many disentanglement methods based on information bottleneck~\cite{autovcqian2019autovc}, perturbation~\cite{speechsplit,nansy}, and representation constraints~\cite{VQMIVC,SIGVC} decompose speech into content, speaker timbre, and speaking style to achieve any-to-any VC, also called zero-shot VC. Based on the recognition-synthesis framework, this work proposes a new method for zero-shot VC.

\begin{figure*}[ht]
\centering
\hspace{-0.5cm}
\begin{minipage}{0.35\linewidth}
\centering
\vspace{1cm}
    \subfigure[Utterance level]{
      \includegraphics[width=1\columnwidth]{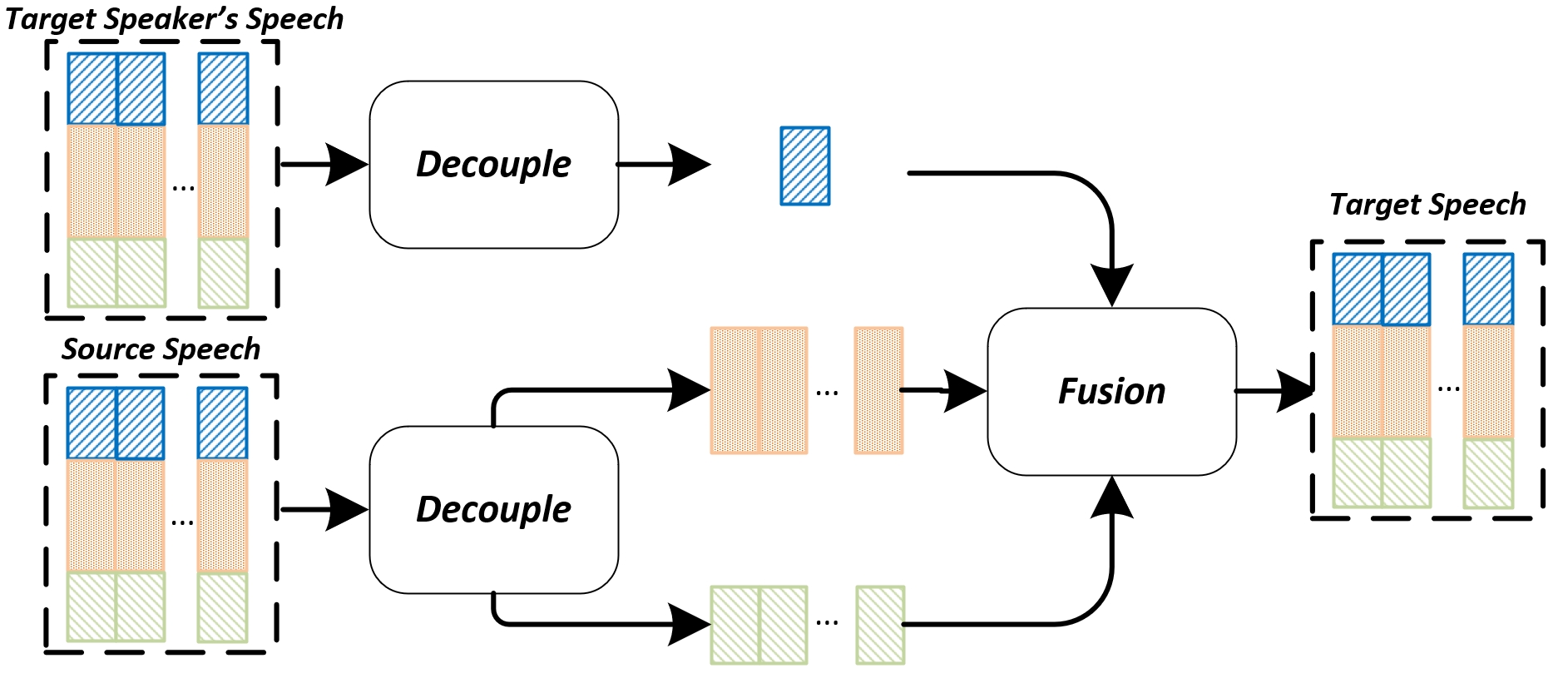}}
\end{minipage}
\begin{minipage}{0.37\linewidth}
\centering
    \subfigure[Fine grained: time varying]{
      \includegraphics[width=1\columnwidth]{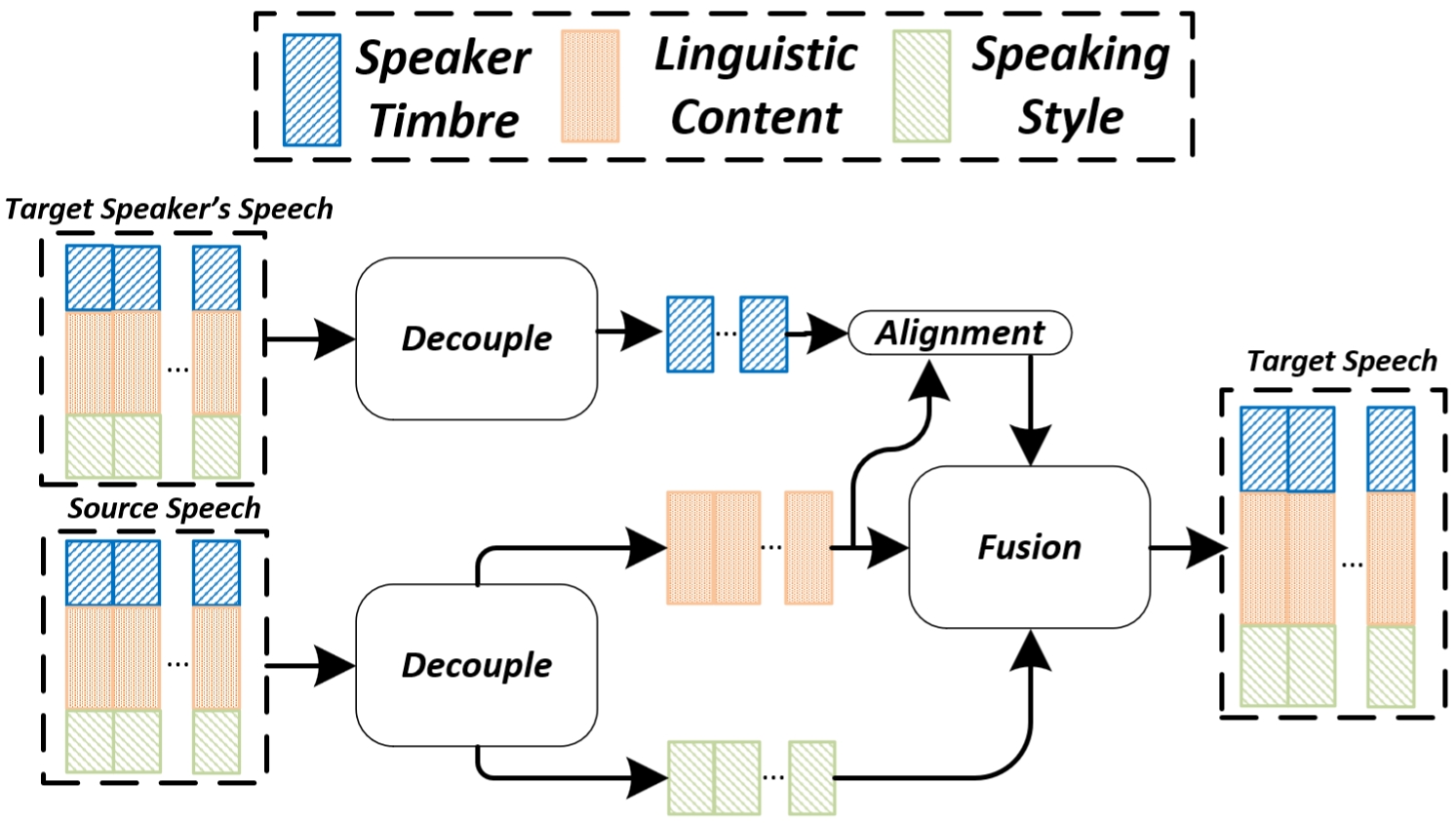}}
\end{minipage}
\begin{minipage}{0.28\linewidth}
\centering
    \subfigure[Fine grained: multi levels]{
      \includegraphics[width=1\columnwidth]{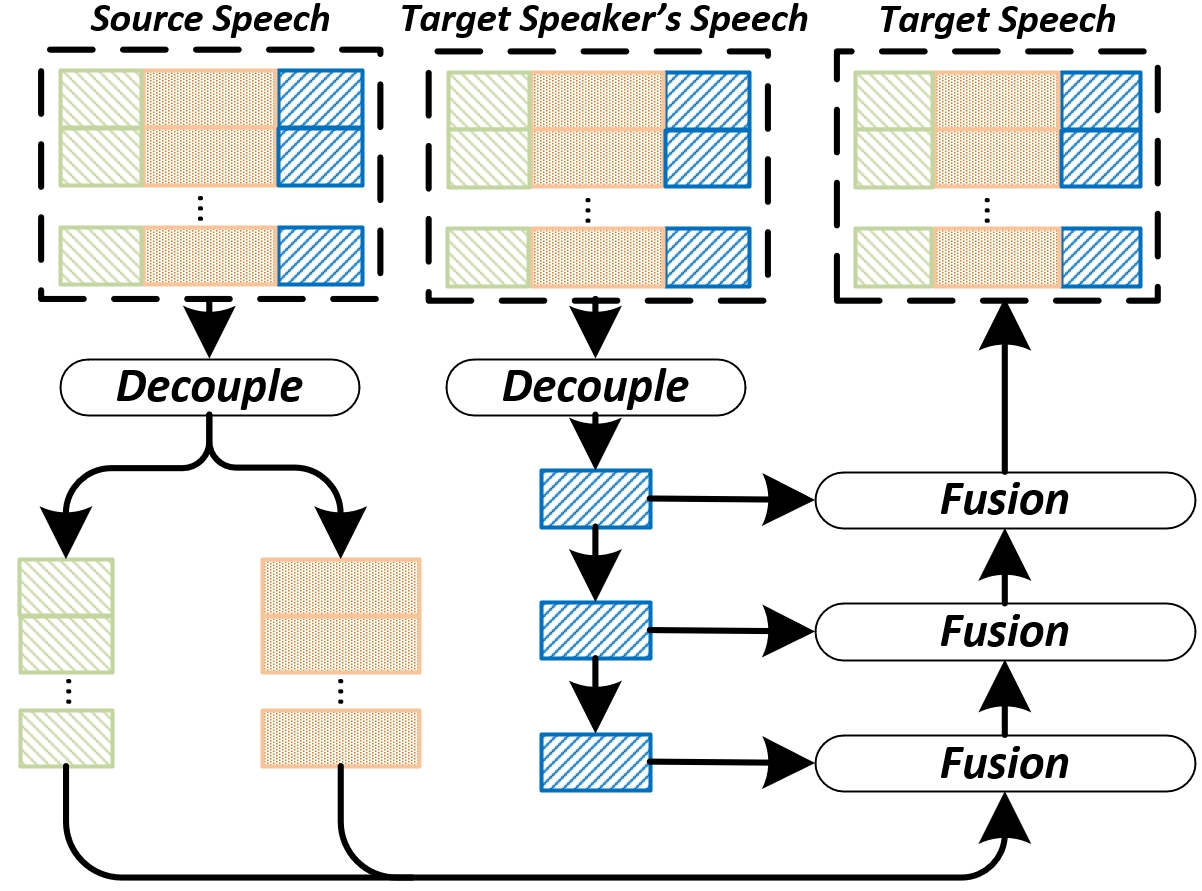}}
\end{minipage}
%\vspace{-10pt}
\caption{The typical frameworks of zero-shot VC. Speaker modeling can be at (a) utterance-level or fine-grained-level with two types: time-varying (b) and multi-level fusion (c).}
\label{fig:zsl_framework}
%\vspace{-17pt}
\end{figure*}

\subsection{Typical Zero-shot VC Framework}
% 如图所示，所有框架的共同点是将语音分解为content speaker 或者 content, style 和 style。在许多工作种，style 并没有被显式的表征出来。在做zero-shot时，从目标说话人语音中提取说话人音色，然后与输入的音频的内容和风格一起生成目标语音。根据不同方法对于说话人音色的建模方式，我们将框架分为如下两种：

The key to voice conversion is to decompose speech into linguistic content and speaker timbre or even speaking style components. Ideally, only the content and style components from the source speech are transferred to the target speech with the target speaker's timbre. 
In many studies~\cite{INchou2019oneshot,AgainVC,FragmentvcAVLin2021,mediumvc}, style is not explicitly considered in the disentanglement. In zero-shot scenarios, the speaker timbre representation is extracted from a short speech utterance of the target speaker, and then the speaker representation is fused with the extracted content and style of the source speech to generate the target speech. The extraction of speaker timbre can be done at utterance level or fine-grained level, as shown in Fig.~\ref{fig:zsl_framework}.

%The VC frameworks can be divided into the following two types depending on the speaker modeling methods.

\subsubsection{Utterance-level Speaker Modeling}
Utterance-level speaker modeling assumes that speaker timbre is static and time-independent within an utterance, as depicted in Figure 1(a). Fixed-length speaker representation is obtained from pre-trained SV models~\cite{autovcqian2019autovc,speakeraware,speechsplit,mediumvc,SIGVC}, or extracted from the target speaker's speech through disentanglement techniques such as instance normalization~\cite{INchou2019oneshot,AgainVC}, adversarial training~\cite{SIGVC}, information bottleneck~\cite{autovcqian2019autovc,VQMIVC,SRDVC}, or perturbation~\cite{speechsplit,nansy}. To integrate the speaker representation with other speech components for speech generation, addition~\cite{avqvc}, concatenation~\cite{autovcqian2019autovc,VQMIVC}, or adaptive instance normalization~\cite{INchou2019oneshot,AgainVC} are commonly used.

% As shown in Fig.~\ref{fig:zsl_framework}(a), in this kind of framework, speaker timbre is assumed to be static and time-independent within an utterance. Fixed-length speaker representation is extracted from pre-trained SV models~\cite{autovcqian2019autovc,speakeraware,speechsplit,mediumvc,SIGVC} or decomposed from the speech of the target speaker by disentanglement, such as instance normalization~\cite{INchou2019oneshot,AgainVC}, adversarial training~\cite{SIGVC}, information bottleneck~\cite{autovcqian2019autovc,VQMIVC,SRDVC}, and perturbation~\cite{speechsplit,nansy}. Besides, addition~\cite{avqvc}, concatenation~\cite{autovcqian2019autovc,VQMIVC}, and adaptive instance normalization~\cite{INchou2019oneshot,AgainVC} are commonly used to fuse speaker representation with other speech components.

\subsubsection{Fine-grained Speaker Modeling}
Fine-grained speaker modeling differs from utterance-level speaker modeling in that it accounts for changes in speaker timbre over time and considers dynamic variation characteristics. Fig.~\ref{fig:zsl_framework}(b) and (c) illustrate two categories of fine-grained speaker modeling: \textit{time-varying} and \textit{multi-level}. In the time-varying speaker modeling~\cite{attbasedzsl,FragmentvcAVLin2021,retriever} as shown in Fig.\ref{fig:zsl_framework}(b), variable-length speaker representation is extracted from the target speaker's speech and then fused with the content and style of the source speech based on the alignment between them. On the other hand, multi-level speaker modeling~\cite{unetts,Xuli_interspeech} shown in Fig.\ref{fig:zsl_framework}(c) typically adopts a \textit{U}-net structure to extract multiple utterance-level speaker representations from stacked layers and then feeds them to the corresponding decoder layers. Recent studies~\cite{FragmentvcAVLin2021,attbasedzsl,retriever} attempt to simultaneously employ time-varying and multi-level speaker modeling. 

Despite this progress, existing methods are insufficient
to capture the dynamic variation characteristic of the speaker
timbre, which varies across temporal and channel regions of speech at different granularities. This insufficient may cause unstable speaker modeling and result in low speaker similarity. Besides, in a time-varying fine-grained speaker modeling framework, speech factors are much easier to entangle with each other, making achieving zero-shot more challenging. This paper focuses on fine-grained speaker modeling in our proposed zero-shot VC framework, which models speaker timbre from both temporal and channel dimensions at different granularities. Comprehensive representations of speech components and a well-designed training strategy ensure the decoupling capabilities of our model.

\subsection{Attention Mechanism in Speaker Modeling}
% attention mechanism in speaker moduling
In recent years, many attention mechanisms have been proposed, such as spatial attention~\cite{RAM}, channel attention~\cite{SEnet}, temporal attention~\cite{TAM}, and their combination~\cite{RSTAN,STA}. Some attention mechanisms have also been explored in speaker modeling. In Desplanques et al.~\cite{ECAPA_TDNN}, Squeeze-and-excitation~\cite{SEnet} block (SE) is introduced to aggregate speaker information along the channel axis. Based on Desplanques et al.~\cite{ECAPA_TDNN}, Liu et al.~\cite{liu2022mfa} introduce multi-scale channel attention to better capture the speaker characteristics at any local frequency region. Besides, Sang et al.\cite{multi-frequency} improve the process of extracting utterance-level speaker representation from frame-level speaker representation by modified channel attention driven by a discrete cosine transform. For zero-shot VC, following Desplanques et al.~\cite{ECAPA_TDNN}, Du et al.~\cite{du21_interspeech}, and Choi et al.~\cite{nansy} utilize SE block to improve the speaker modeling ability for zero-shot VC. CA-VC uses channel attention to improve content learning and encourage utterance-level speaker modeling. Besides, Retriever~\cite{retriever} uses cross-attention~\cite{cross-attention} to learn a set of permutation invariant tokens to represent speaker timbre. In variable-length speaker modeling~\cite{attbasedzsl,retriever,FragmentvcAVLin2021}, scale-dot attention and cross-attention are usually utilized to fuse speaker and content representations. Previous studies~\cite{du21_interspeech,nansy,CAVC} have explored the correlation between channel and speaker timbre information in zero-shot VC, but only at the utterance level and single channel dimension. In contrast, our proposed method focuses on fine-grained speaker modeling and performs speaker retrieval in both temporal and channel dimensions.

\begin{figure}[h]	
\centering
%\vspace{-1pt}
\includegraphics[width=1\linewidth]{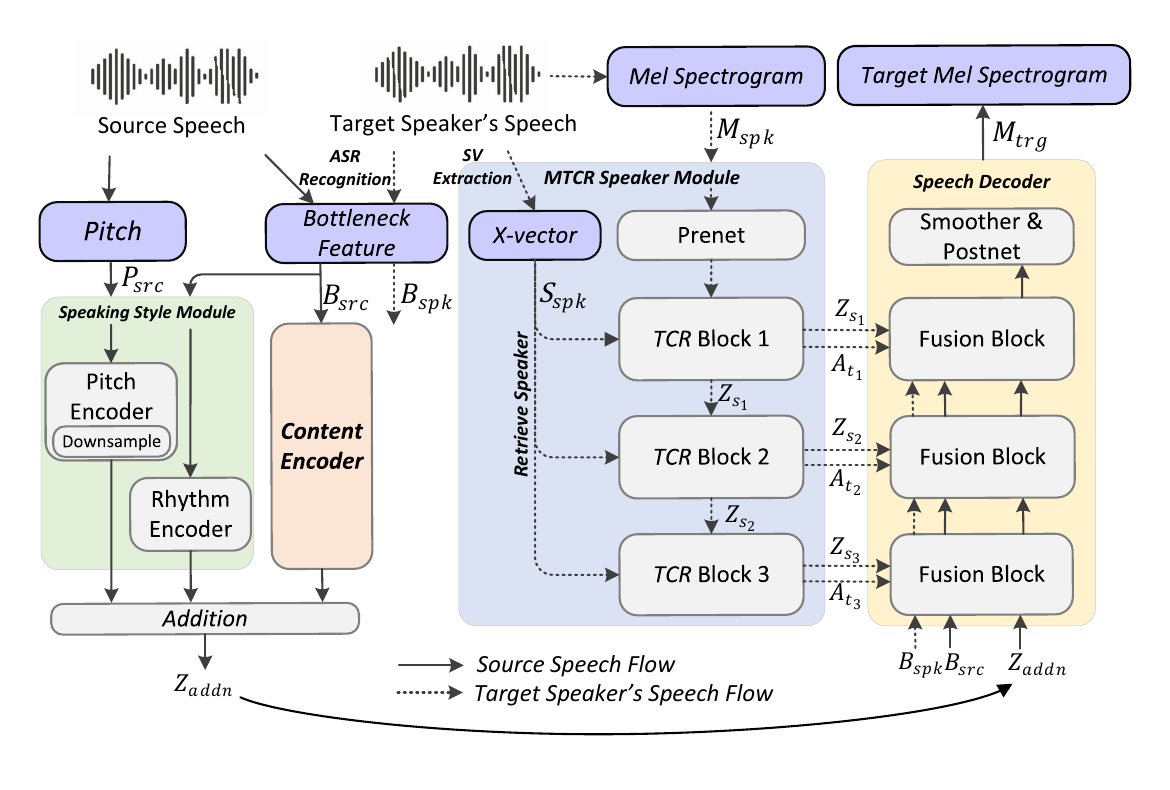}
\caption{The framework of the proposed MTCR-VC model. }
\label{fig:MTCR-VC}
%\vspace{-10pt}
\end{figure}

\section{METHODOLOGY}
\label{sc:method}
\subsection{The Framework of MTCR-VC}
 %\vspace{-3pt}

In this paper, we introduce multi-level temporal channel retrieval-based voice conversion (MTCR-VC), a novel approach that integrates time-varying and multi-level speaker modeling frameworks with substantial improvements, as depicted in Figure~\ref{fig:MTCR-VC}. Based on the structure of FragmentVC~\cite{FragmentvcAVLin2021}, MTCR-VC disentangles speech into speaking style, linguistic content, and speaker timbre using a speaking style module, a content encoder, and an MTCR speaker module, respectively. The speaking style and linguistic content representations are learned from the source speech's pitch contour $P_{src}$ and BNF $B_{src}$. Then the content, pitch, and rhythm representations are added to form $Z_{add}$. The MTCR speaker module extracts multiple fine-grained speaker representations from the target speaker's speech. Finally, the speech decoder progressively combines the learned speech representations to produce the corresponding target mel spectrogram $M_{trg}$, which conveys the target speaker's timbre while maintaining the source speech's speaking style and linguistic content.

\subsubsection{Content Encoder}
The content encoder is a Conformer-like structure that learns the linguistic content representation, as indicated in the orange section on the left of Fig.~\ref{fig:MTCR-VC}. 
Here, the bottleneck features (BNF) serve as the input for the content encoder, as BNF is considered as a speaker-irrelevant linguistic representation~\cite{PPGSun2016PhoneticPF,wang2022delivering}, obtained from the encoder of a well-trained ASR model.

% comprehensive characterizations of the speech components 
% \subsubsection{Speaking Style Module}
\subsubsection{Speaking Style Module}
As shown in the green part on the left of Fig.~\ref{fig:MTCR-VC}, the speaking style module comprises a pitch encoder and a rhythm encoder that aim to learn pitch and rhythm representations, respectively. Specifically, the pitch encoder consists of a fully connected (FC) layer to learn the pitch representation from the normalized pitch contour $P_{src}$ of the source speech, followed by a downsample layer that reduces speaker-related information. On the other hand, the rhythm encoder, which has a similar structure to the reference encoder in~\cite{ReferenceSkerryRyan2018Reference}, is adapted to extract the utterance-level rhythm representation from the BNF $B_{src}$ of the source speech. Note that BNF is considered to be a speaker-irrelevant linguistic feature and is widely used in VC. Recent studies~\cite{wang21g_interspeech,cross,msmvc} use BNF for rhythm modeling and also prove the BNF retains rhythm information.

% but rhythm-related~\cite{wang21g_interspeech,wang2022delivering}, a speaker-irrelevant rhythm representation can be obtained from it.

%  
% . And a down-sample layer is also adopted to further reduce the speaker-related information from the source speech. 
%and reduces the speaker-related information in the source speech, yielding the pitch representation. 

%\subsubsection{Content Encoder}

% Instead of jointly learning content representation from speech~\cite{autovcqian2019autovc,INchou2019oneshot,VQMIVC}, the content encoder uses BNF of the source speech as the input. 

 % Note that the speaking style module~\cite{wang21g_interspeech} and the content encoder~\cite{wang2022delivering} have been adopted in previous studies and achieved good performance. The proposed MTCR speaker module and training strategy are discussed in Section~\ref{sc:method}. To accommodate the new speaker modeling, we make some small modifications to the speech decoder~\cite{FragmentvcAVLin2021} and also describe it in Section~\ref{sc:method}.
%It is worth noting that previous studies have successfully adopted the speaking style module~\cite{wang21g_interspeech} and content encoder~\cite{wang2022delivering}, which achieved good performance. 

Based on the above structure, we propose the MTCR speaker module and a training strategy, which are elaborated in Section~\ref{sec:retri} and \ref{sc:training}. Moreover, we modify the speech decoder~\cite{FragmentvcAVLin2021} to facilitate the integration of the MTCR speaker module, which are also detailed in Section~\ref{sc:decoder}.

%To facilitate the integration of the MTCR speaker module, we made minor modifications to the speech decoder~\cite{FragmentvcAVLin2021}, which are also detailed in Section~\ref{sc:method}.

\begin{figure*}[ht]	
\centering
%\vspace{-1pt}
\includegraphics[width=1\linewidth]{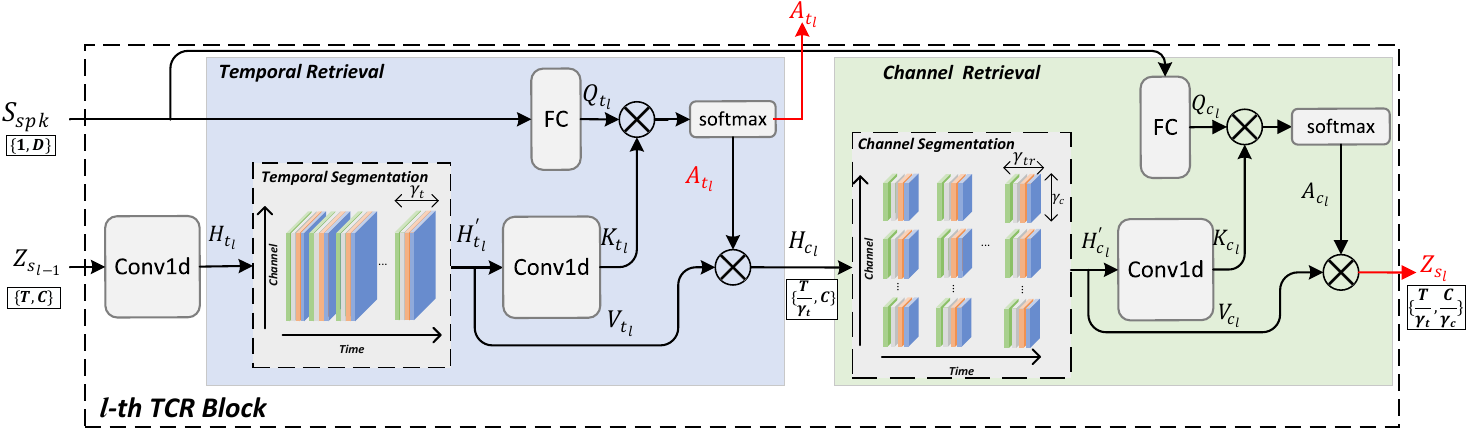}
\caption{The architecture of the TCR block. Note that the symbols in red color represent the outputs of the block.}
\label{fig:TCR}
%\vspace{-10pt}
\end{figure*}

% \section{Multi-level Temporal-channel Speaker Retrieval}
% \label{sc:method}
%  % %\vspace{-5pt}
% This section details the MTCR speaker module and the cycle-based training strategy to facilitate the training of MTCR-VC. Additionally, the speech decoder, adjusted to the proposed speaker modeling, is also discussed in this section.

\subsection{MTCR Speaker Module: Speaker Modeling}
% \subsection{MTCR Speaker Module: Multi-level Temporal-channel Retrieval}
% 添加对于多层的描述
\label{sec:retri}

% Speaker timbre is a dynamic characteristic that varies across temporal and channel regions of speech, owing to the changes in the speech signal over time and frequency~\cite{audio,attbasedzsl,liu2022mfa}. Moreover, the hierarchical nature of the human speech production mechanism results in speaker timbre variation at different granularities. Inspired by the representation learning ability of attention mechanism~\cite{liu2022mfa,multi-frequency,ECAPA_TDNN,RSTAN,STA}, previous studies~\cite{du21_interspeech,nansy,CAVC} have explored the correlation between channel and speaker timbre information in zero-shot VC, but only at the utterance-level and single channel dimension. In this work, we propose the \textbf{m}ulti-level \textbf{t}emporal-\textbf{c}hannel \textbf{r}etrieval (MTCR) speaker module for retrieving speaker-related information from the target speaker's speech in both temporal and channel regions with different granularities. Specifically, as shown in the blue part of Fig.~\ref{fig:MTCR-VC}, the MTCR speaker module comprises multiple TCR blocks to achieve multi-granularity speaker retrieval. Each TCR block sequentially performs temporal and channel retrieval at a specific granularity. And the speaker embedding (x-vector) derived from a SOTA SV model~\cite{ECAPA_TDNN} is considered an ideal query to guide the speaker retrieval since the SV model~\cite{ECAPA_TDNN} achieves high accuracy in speaker verification and learns accurate speaker discriminative ability~\cite{ECAPA_TDNN, SIGVC, mediumvc}.

Speaker timbre is a dynamic characteristic that varies across temporal and channel regions of speech at different granularities. Inspired by the representation learning ability of attention mechanism~\cite{liu2022mfa,RSTAN}, we propose the \textbf{m}ulti-level \textbf{t}emporal-\textbf{c}hannel \textbf{r}etrieval (MTCR) speaker module for retrieving speaker-related information from the target speaker's speech in both temporal and channel regions with different granularities. Specifically, as shown in the blue part of Fig.~\ref{fig:MTCR-VC}, the MTCR speaker module comprises multiple TCR blocks to achieve multi-granularity speaker retrieval. Each TCR block sequentially performs temporal and channel retrieval at a specific granularity. The speaker embedding (x-vector) derived from an SV model~\cite{ECAPA_TDNN} is considered an ideal query to guide the speaker retrieval since the SV model achieves high accuracy in speaker verification and learns accurate speaker discriminative ability.

As shown in Fig.~\ref{fig:TCR}, the TCR block comprises a convolution layer, a temporal retrieval module, and a channel retrieval module. The attention mechanism is used in both retrievals to aggregate speaker information in the temporal and channel dimensions by a weighted sum. Specifically, it uses x-vector $S_{spk} \in \mathbb{R}^{1 \times D}$ as \textit{query} and the previous block's output $Z_{s_{l-1}} \in \mathbb{R}^{T \times C}$ as \textit{key} and \textit{value} to hierarchically performs temporal and channel retrieval, where $l$ represents the index of the TCR block, $T$ represents the temporal length, $C$ represents the channel length, and $D$ represents the x-vector's dimension. 
% In the following, we describe the temporal and channel retrieval in detail. 

% After passing $Z_{s_{l-1}}$ into the convolution layer, the hidden feature $H_{t_{l}} \in \mathbb{R}^{T \times C}$ is generated. Finally, temporal and channel retrieval are performed and output $A_{t_{l}}$ and $Z_{s_{l}}$.
% the results of temporal and channel retrieval will be inferred and form the output of the block. The whole structure and process can be seen in Fig.~\ref{fig:STCR-Fusion}c.

% %\vspace{-3pt}
\textbf{Temporal Retrieval.} The objective of our work is to enable the speaker module to know \textit{when} it should pay more attention to capture speaker information. For this purpose, we adopt temporal retrieval. As speech varies in duration and speaker timbre exhibits different characteristics at various temporal granularities, temporal segmentation, in which the features are equally divided into several groups along the temporal dimension, must be considered before applying the attention mechanism. Specifically, after the convolution layer processing the input $Z_{s_{l-1}}$ into the output $H_{t_{l}}$, we perform temporal segmentation on $H_{t_{l}}$ using a temporal scale factor $\gamma_{t}$. Then the resulting segmentation $H^{'}_{t_{l}} \in \mathbb{R}^{\frac{T}{\gamma_{t}} \times \gamma_{t} \times C}$ is used as the key $K_{t_{l}}$ and value $V_{t_{l}}$. Meanwhile, the x-vector $S_{spk}$ is transformed by an FC layer and used as the query $Q_{t_{l}} \in \mathbb{R}^{1 \times C}$. 
%In other words, we utilize the discriminative attribute of the target speaker to retrieve the correlated information of speaker timbre along the temporal axis of speech. 
By measuring the similarity between $Q_{t_{l}}$ and $K_{t_{l}}$, we obtain an attention map $A_{t_{l}} \in \mathbb{R}^{\frac{T}{\gamma_{t}} \times 1 \times \gamma_{t}}$ that captures the distribution of speaker timbre in each temporal segment. Finally, we multiply $A_{t_{l}}$ and $V_{t_{l}}$, yielding $H_{c_{l}} \in \mathbb{R}^{\frac{T}{\gamma_{t}} \times C}$ that we also use for the following channel retrieval. The entire process of temporal retrieval can be summarized as follows:
% %\vspace{-5pt}
\begin{equation}
    V_{t_{l}} = H^{'}_{t_{l}} = Segmentation(H_{t_{l}})
%\vspace{-7pt}
\end{equation}
\begin{equation}
    K_{t_{l}} = Conv(H^{'}_{t_{l}})
\end{equation}
\begin{equation}
    Q_{t_{l}} = FC(S_{spk})
%\vspace{-7pt}
\end{equation}
\begin{equation}
    H_{c_{l}}=A_{t_{l}}V_{t_{l}}=softmax(\frac{Q_{t_{l}} K^\top_{t_{l}}}{\sqrt{C}})V_{t_{l}}
\end{equation}

% 在我们得工作中，我们希望再说话人建模得过程中模型能够更加清楚得知道什么时候需要关注时序序列。为了实现这个目的，我们提出了时序检索。在有卷积层总结之后，由于语音是不定长的同时说话人特性在不同的时间分辨率下有不同的特性，根剧比例因子  对hidden feature 进行时序切片。所以特征维度从【】变成怕【】.切片的结果被分别处理作为了key和value. 在FC层的维度转换之后，Strg被用作Query去指示说话人的音色特性。换句话说，我们使用目标说话的判别性属性去衡量什么时间说话人音色信息产生的更多。通过query和key之间的相似度衡量，attention map A 表明了在每个序列中说话人信息的时序分布 引导了时序查询最终得到H\.整个过程可以表述维：

\textbf{Channel Retrieval.} To retrieve speaker timbre from speech, it is crucial to not only determine \textit{when} the speaker timbre is present but also \textit{where} it exists in the channel dimension. Therefore, we use both temporal and channel retrieval techniques to achieve this.
% Except for knowing \textit{when} is more related to the speaker timbre, it is also essential to find \textit{where} in the channel dimension reflects more speaker timbre information. Hence, we utilize both temporal and channel retrieval to achieve this end. 
Fig.~\ref{fig:TCR} illustrates our approach, which begins with channel segmentation applied to $H_{c_{l}}$. Since the speaker timbre changes with
 time-varying style or content, the distribution of speaker timbre on the channel is also related to the temporal changes. Therefore, we consider each channel segment's temporal range $\gamma_{tr}$ during the channel segmentation.
% As shown in Fig.~\ref{fig:TCR}, similar to the temporal retrieval, we first apply channel segmentation to $H_{c_{l}}$. Since the speaker timbre dynamically changes over time-varying speaking styles or speech content, the distribution of speaker timbre on the channel is also related to the temporal dimension. 
% For example, under time-varying speaking styles or speech content, the changes of speaker characteristics in different channels are also reflected in the  temporal dimension. 
% Thus, it is necessary to consider the temporal range $\gamma_{tr}$ of each channel segment when doing channel segmentation. 
After the channel segmentation, the result $H^{'}_{c_{l}}$ is in the shape of $\{\frac{T^{'}}{\gamma_{tr}},\frac{C}{\gamma_{c}},\gamma_{c},\gamma_{tr}\}$, where $T^{'}$ represents $\frac{T}{\gamma_{t}}$ mentioned above, and $\gamma_{c}$ represents the channel scale factor. Then, $H^{'}_{c_{l}}$ is used as the key $K_{c_{l}}$ and value $V_{c_{l}}$. Meanwhile, after processing by an FC layer, the x-vector $S_{spk}$ is used as the query $Q_{c_{l}}$. To compute the attention map $A_{c_{l}} \in \mathbb{R}^{ \frac{T^{'}}{\gamma_{tr}} \times \frac{C}{\gamma_{c}} \times 1 \times \gamma_{c}}$ for each channel segment, we measure the similarity between $K_{c_{l}}$ and $Q_{c_{l}}$. Finally, we obtain the TCR block result $Z_{s_{l}}$, with dimensions of $\{\frac{T}{\gamma_{t}},\frac{C}{\gamma_{c}}\}$, which is computed by multiplying $A_{c_{l}}$ and $V_{c_{l}}$.

% Hence, attention map $A_{c_{l}} \in \mathbb{R}^{ \frac{T^{'}}{\gamma_{tl}} \times \frac{C}{\gamma_{c}} \times 1 \times \gamma_{c}}$ of each channel segment can be computed by measuring the similarity between $K_{c_{l}}$ and $Q_{c_{l}}$. Finally, $Z_{s_{l}}$ with the dimension of $\{\frac{T}{\gamma_{t}},\frac{C}{\gamma_{c}}\}$ computed by multiplying $A_{c_{l}}$ and $V_{c_{l}}$ is used as the output of the TCR block. 
It should be noted that the scale factors $\gamma_{t}$ and $\gamma_{c}$ in segmentation determine the receptive region of the attention mechanism and affect the granularity of temporal and channel regions for speaker retrieval. Additionally,  the segmentation process within each TCR block relies on the segmentation outcome of the previous block, which means that the granularity of speaker retrieval increases with the deepening of the layer of the TCR block.

\begin{figure}[h]	
\centering
%\vspace{-1pt}
% \includegraphics[width=0.7\linewidth]{Figure/fusing.pdf}
\begin{minipage}{0.5\linewidth}
\centering
    \subfigure[Fusing block]{
      \includegraphics[width=1\columnwidth]{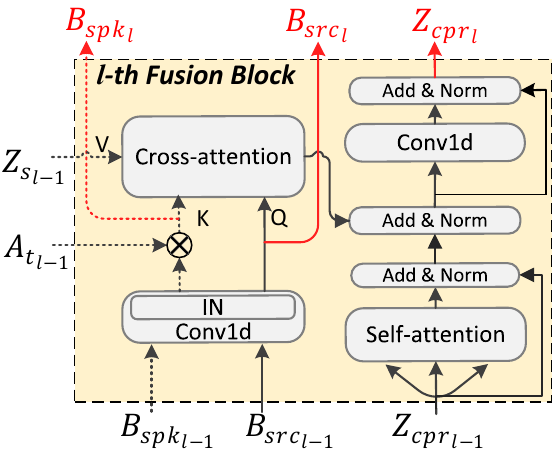}}
\end{minipage}
\begin{minipage}{0.45\linewidth}
\centering
    \subfigure[Extractor~\cite{FragmentvcAVLin2021}]{
      \includegraphics[width=0.85\columnwidth]{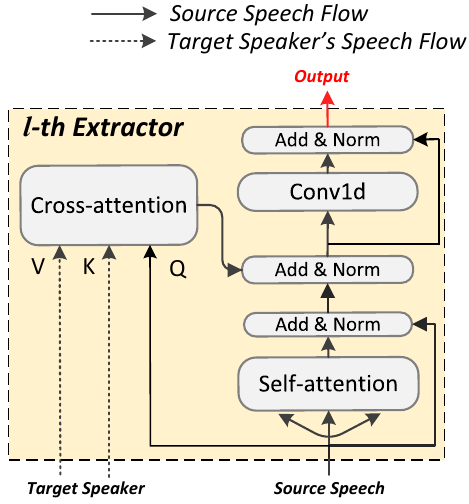}}
\end{minipage}

\caption{The (a) fusion block in the speech decoder. The symbols in red color represent the outputs of the fusion block. The architecture of the fusion block follows the design of Extractor~\cite{FragmentvcAVLin2021}.}
\label{fig:Fusion}
%\vspace{-10pt}
\end{figure}

\subsection{Speech Decoder: Speech Representation Fusion}
\label{sc:decoder}
%\subsection{Speech Decoder: Fusion and Generation}
To accommodate the proposed speaker modeling, our framework adopts a speech decoder, modified from the structure in a previous study~\cite{FragmentvcAVLin2021}. The speech decoder, shown in yellow in Figure~\ref{fig:MTCR-VC}, comprises multi-level fusion blocks, a smoother, and a postnet. The multi-level fusion blocks progressively combine the representations of speaking style, linguistic content, and speaker timbre, while the smoother and postnet generate the target speech spectrogram using the final fusion block's output. Specifically, the fusion block, as illustrated in Figure~\ref{fig:Fusion}, is similar to the extractor in FragmentVC~\cite{FragmentvcAVLin2021}, with the cross-attention~\cite{vaswani2017attention} generating content-based alignment between the source speech and the target speaker's speech. In contrast to the original extractor, the cross-attention uses the speaker-independent BNFs $B_{src}$ and $B_{spk}$ extracted from the source speech and target speaker's speech as the query and key, respectively. This modification ensures that the speaker timbre does not influence the content-based alignment. Furthermore, the attention map of temporal retrieval $A_{t_{l-1}}$ maintains the same sequence lengths of the key and value $Z_{s_{l}}$.

\begin{figure}[h]	
\centering
\begin{minipage}{0.34\linewidth}
    \subfigure[Paired path]{
      \includegraphics[width=1\columnwidth]{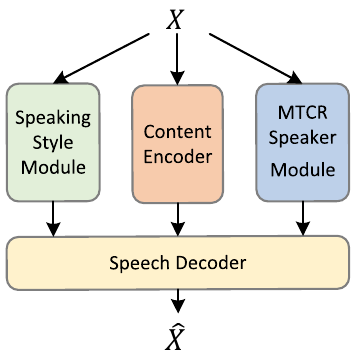}}
\end{minipage}
\begin{minipage}{0.63\linewidth}
    \subfigure[Unpaired path]{
      \includegraphics[width=1\columnwidth]{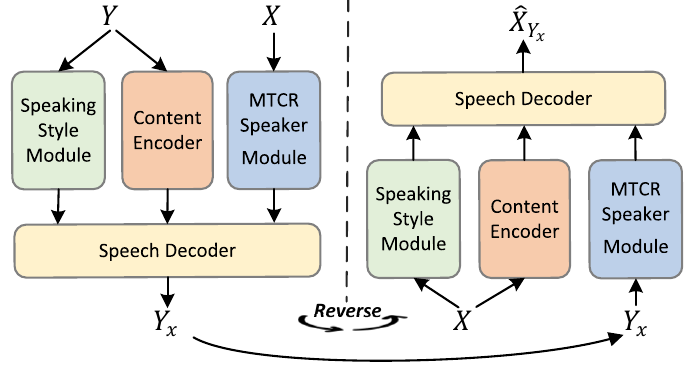}}
\end{minipage}\\

% \centering
% \begin{minipage}{1\linewidth}
% \centering
%     \subfigure[Perceptual Constraints]{
%       \includegraphics[width=0.9\columnwidth]{Figure/training_loss.pdf}}
% \end{minipage}
\caption{The cycle-based training process of MTCR-VC. (a) Paired Path. (b) Unpaired Path. Note that speech utterances X and Y come from different speakers with different linguistic content.}
\label{fig:training}
%\vspace{-10pt}
\end{figure}
% The block in purple represents pre-trained model, which is fixed without parameter update in the VC model.
%\vspace{-5pt}

\subsection{Cycle-based Training Strategy}
\label{sc:training}
%\vspace{-3pt}
Generally, during the training process of a VC model, the style, content, and speaker representations are usually learned from the same speech utterance. The converted speech is the paired reconstruction result. However, this process can result in the entanglement of speech representations, particularly in fine-grained speaker modeling~\cite{FragmentvcAVLin2021,attbasedzsl}. To address this issue, previous studies~\cite{FragmentvcAVLin2021,attbasedzsl} use two different speech utterances from the same speaker to extract different speech representations, with one utterance used to extract speaker representation and the other used to extract content and style representations. Nonetheless, when performing zero-shot VC, style, content, and speaker timbre come from different speakers' speech utterances, leading to the unpaired reconstruction. This mismatch between paired training and unpaired zero-shot VC inference can cause insufficient speech representation disentanglement, as paired training alone cannot ensure the disentanglement ability between different speakers. To better facilitate speech disentanglement and model generalization, following the cycle-consistency idea of CycleGAN~\cite{maskcyclegan}, we introduce a cycle-based training strategy with paired and unpaired paths in which two different speech utterances $X$ and $Y$ from two speakers are used to learn different representations during the training process, as presented in Fig.~\ref{fig:training}.

 % Generally, in the training process of VC model, the representations of style, content, and speaker are learned from the same speech, and the converted speech is the paired reconstruction result. But it will lead to speech representations entanglement in fine-grained speaker modeling~\cite{FragmentvcAVLin2021,attbasedzsl}. Previous studies~\cite{FragmentvcAVLin2021,s2vc} usually use two different speech recordings from the same speaker to extract different speech representations, e.g., one is used to extract speaker representation, and the other is used to extract content and style representations. But when performing zero-shot VC, style, content, and speaker timbre come from different speakers' speech recordings, leading to the unpaired reconstruction.The mismatch between paired training and unpaired zero-shot VC inference causes insufficient disentanglement since paired training cannot ensure the disentanglement ability between different speakers. To facilitate speech disentanglement and model generalization, inspired by the training process of CycleGAN~\cite{maskcyclegan}, a cycle-based training strategy with paired and unpaired paths is designed, in which two different speech recordings from two speakers involve learning different representations in the training process.

\textbf{Paired Path.} To ensure the quality of the reconstructed speech, we employ the paired path, as depicted by the blue arrow in Fig.~\ref{fig:training}(a). In the paired path, the speaking style module, content encoder, and MTCR speaker module take the same speech utterance $X$ as input to extract style, content, and speaker representations, respectively. The speech decoder then combines these representations to produce the reconstructed speech utterance $\hat{X}$. The reconstruction quality is evaluated using the mel loss $L_{mel}^{\hat{X}}$, which measures the mean square error (MSE) between $X$ and $\hat{X}$.

\textbf{Unpaired Path.} The brown arrow in Fig.~\ref{fig:training}(a) illustrates the unpaired path in which speech utterances $X$ and $Y$ from different speakers are utilized to generate the converted result $Y_{x}$ of zero-shot VC and then re-convert it to $\hat{X}_{Y_{x}}$. Initially, the model produces $Y_x$ that retains the speaker timbre of $X$ but the speaking style and linguistic content of $Y$. Then, $Y_x$ is employed to extract the speaker timbre and recover $X$ with the linguistic content and speaking style of $X$. In other words, the unpaired path ensures
that the input speech is not only converted to the speech of any speaker but also reversibly transformed back to the
original speaker’s speech, seeking optimal disentanglement and maintaining high fidelity.

\begin{figure}[h]	
\centering
%\vspace{-1pt}
\includegraphics[width=1\linewidth]{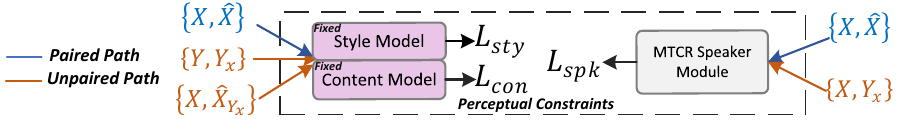}
\caption{The perceptual constraints used in the training of MTCR-VC. The block in purple represents the pre-trained model, which is fixed without parameter update in the model.}
\label{fig:perceptual}
%\vspace{-10pt}
\end{figure}

Since the corresponding ground-truth speech of $Y_x$ is unavailable in the unpaired path (Fig~\ref{fig:training}(b)), only mel reconstruction loss is hard to constrain the training process. We introduce perceptual constraints in style,
content, and speaker to supervise the training process and improve the disentanglement ability. The perceptual constraints are illustrated in Fig.~\ref{fig:perceptual}. Similarly to Wang et al.~\cite{wang2022delivering}, we calculate the content loss $\mathcal{L}_{con}$ and style loss $\mathcal{L}_{sty}$ to ensure the consistency of content and style between the converted speech and the source speech. The speaker loss $\mathcal{L}_{spk}$ calculated via the MTCR speaker module measures the global difference of speaker representations between the converted speech and the target speaker's speech at each layer. The details of the loss functions are introduced in Section~\ref{sec:loss}.

\begin{figure}[h]	
\centering
%\vspace{-1pt}
\includegraphics[width=0.8\linewidth]{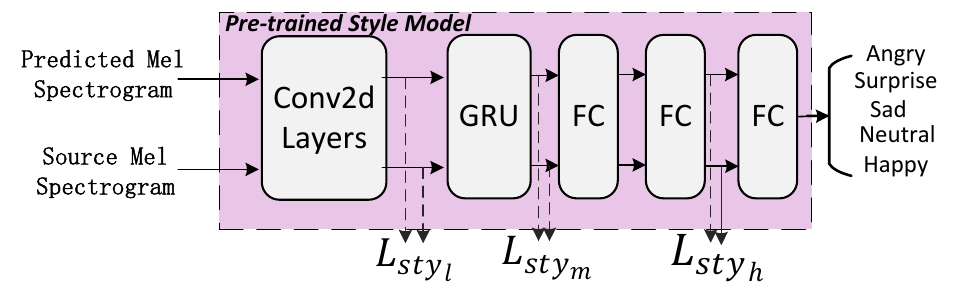}
\caption{The architecture of the style model.}
\label{fig:style}
%\vspace{-10pt}
\end{figure}

\subsection{Overall Loss Functions}
\label{sec:loss}

\subsubsection{Mel loss}
To effectively optimize the proposed model, mel loss $L_{mel}$ is usually introduced to ensure the reconstruction quality of the mel spectrogram. L2 distance between predicted mel spectrogram $\hat{M}$ and ground-truth mel spectrogram $M$ is adopted as the mel loss, which is defined as:
\begin{equation}
    \mathcal{L}_{mel} = ||M-\hat{M}||^{2}_{2}.
\end{equation}

% With only speaker timbre constraint forcing the learning of new speaker timbre during adaptation, 
% the adapted model would be over-fitting to the limited utterances. It is necessary to prevent the abilities of the model learned in base model training from being forgotten, including style delivery, content preservation, and high-quality generation. The regularization constraints on these aspects are employed to achieve the purpose and make the model parameters focus on being optimized in the speaker timbre space during adaptation.

% \subsubsection{Speaking style consistency}
\subsubsection{Style loss} 
As shown in Fig.~\ref{fig:style}, a pre-trained speech emotion recognition (SER) model~\cite{liurui2021expressive} is used as a style model to calculate the style loss. $\mathcal{L}_{sty}$ is calculated from different levels, accounting for the abstraction degree of the hidden representation to the style. Features $H_{l}$, $H_{m}$, and $H_{h}$ from low- to high-level hidden layers, respectively, are used to obtain the style loss:
\begin{equation}
% \footnotesize
    \mathcal{L}_{sty_{s}}=||H_s-\hat{H_s}||^{2}_{2}, s\in \{ l, m, h \}
\end{equation}
where $h$ and $\hat{h}$ are obtained from the source mel spectrogram and predicted mel spectrogram, respectively.

% \subsubsection{Linguistic content}

\subsubsection{Content loss} 
The content loss $\mathcal{L}_{con}$ is generated by a pre-trained content model that has the same architecture as the CBHG module~\cite{TacotronWang2017}. The content model is trained for BNF prediction. During the training of MTCR-VC, the content loss between BNF features $B$ and $\hat{B}$ extracted from the source mel spectrogram and predicted mel spectrogram can be described as:
\begin{equation}
% \footnotesize
\mathcal{L}_{con}=||B-\hat{B}||^{2}_{2}
\end{equation}

% \subsubsection{Speech quality}
\subsubsection{Speaker loss} 
To ensure the consistency of speaker-related information after conversion, the MTCR speaker module is used to calculate the speaker loss $\mathcal{L}_{spk}$ from different levels. Since the speaker representations $Z_{s_l}$ from different speech might have variable length, $\mathcal{L}_{spk}$ only measure the global difference, which can be described as:
\begin{equation}
% \footnotesize
    \mathcal{L}_{spk_{l}}=||Avg(Z_{s_l})-Avg(\hat{Z_{s_l}})||^{2}_{2}, l\in \{ 1, 2, 3 \}
\end{equation}
where $Z_{s_l}$ and $\hat{Z_{s_l}}$ are obtained from the target speaker's mel spectrogram and predicted mel spectrogram, respectively. $Avg(\cdot)$ means averaging the representation along the temporal axis.

 % These two constraints ensure the consistency of content and style between the converted speech and the source speech. The speaker constraint $\mathcal{L}_{spk}$ calculated via the MTCR speaker module measures the global difference of speaker representations between the converted speech and the target speaker's speech at each layer. It is worth noting that the speaker representation involved in calculating $\mathcal{L}_{spk}$ is first averaged along the temporal axis. 

 As shown in Fig.~\ref{fig:perceptual}, all three perceptual constraints are applied to the converted result $Y_{x}$ and $\hat{X}$, while only content and style constraints are used for the re-converted result $\hat{X}_{Y_{x}}$ since its corresponding target speaker's speech is also the converted speech. The training objectives of the paired path $\mathcal{L}_{pair}$ and unpaired paths $\mathcal{L}_{unpair}$ can be described as:
 \begin{equation}
\begin{split}
\mathcal{L}_{pair}=\mathcal{L}_{mel}^{\hat{x}}+\lambda_{spk}\mathcal{L}^{\hat{x}}_{spk}+\lambda_{sty}\mathcal{L}^{\hat{x}}_{sty}+\lambda_{con}\mathcal{L}^{\hat{x}}_{con}
\end{split}
\end{equation}
\begin{equation}
%\vspace{-5pt}
\begin{split}
\mathcal{L}_{unpair}=\lambda_{mel}\mathcal{L}_{mel}^{\hat{X}_{Y_{x}}}+\lambda_{spk}\mathcal{L}_{spk}^{Y_{x}}+\lambda_{sty}\\(\mathcal{L}_{sty}^{Y_{x}}+\mathcal{L}_{sty}^{\hat{X}_{Y_{x}}})+\lambda_{con}(\mathcal{L}_{con}^{Y_{x}}+\mathcal{L}_{con}^{\hat{X}_{Y_{x}}})
\end{split}
\end{equation}
where $\lambda$ before each loss term represents the corresponding weight. 
%Note that $\hat{X}_{Y_{x}}$ does not involve in calculating speaker constraint because the corresponding speaker speech $Y_{x}$ is also the converted result. 
In our work, the two paired and unpaired paths are performed simultaneously, so the total loss $\mathcal{L}_{total}$ of the proposed model is $\mathcal{L}_{total}=\mathcal{L}_{pair}+\mathcal{L}_{unpair}$.

\section{Experimental Setup}
\label{sc:experiments}

%To evaluate the performance of the proposed MTCR-VC on the zero-shot VC task, experiments are conducted on a multi-speaker English speech corpus. 
In this section, we first introduce the database used in our work. Next, we provide the implementation details of the proposed model. Finally, we present comparison and evaluation metrics used in experiments.

\subsection{Datasets}
During the training stage, 1,000 speakers from the English dataset LibriTTS~\cite{LibriTTS} are used to train the VC model.
% In the zero-shot testing stage, we randomly select 12 target speakers, including 6 pre-reserved in-dataset (ID) speakers from LibriTTS and 6 out-of-dataset (OOD) speakers from VCTK~\cite{VCTK}, CMU Arctic~\cite{CMU-Arctic}, and HiFi-TTS dataset~\cite{hifi} (2 speakers for each dataset).
In the zero-shot voice conversion stage, we test two types of target speakers, including pre-reserved in-dataset (ID) speakers from LibriTTS and out-of-dataset (OOD) speakers from VCTK~\cite{VCTK}, CMU Arctic~\cite{CMU-Arctic}, and HiFi-TTS datasets~\cite{hifi}. The OOD testing is particularly used to show the robustness of our approach. Information regarding the recording domain, environment, and signal-to-noise ratio (SNR) of the datasets used in this work can be found in Table~\ref{exp:datatset}. Additionally, speech utterances from VCTK and CMU Arctic are selected as source speech for the test. We randomly sampled 1,000 testing pairs for ID and OOD speakers. We use the open-source conformer ASR model\footnote{https://github.com/wenet-e2e/wenet}, trained on LibriSpeech~\cite{librispeech}, to extract BNF. The SV model~\cite{ECAPA_TDNN} is trained on Voxceleb2~\cite{voxceleb2}. The style and content models for perceptual constraints are trained on ESD~\cite{zhou2021seenESD} and LibriTTS, respectively. To reconstruct waveform from mel spectrogram, we utilize a universal vocoder TFGAN~\cite{tian2020tfgan}, which is trained on 1,000 hours of an internal dataset.

\begin{table}[ht]
\centering
%\scriptsize
\setlength{\tabcolsep}{0.8mm}
\renewcommand\arraystretch{2}
\caption{The information of the datasets used in experiments. }
% \caption{Validation of each level TCR block.}
%\vspace{-5pt}
\label{exp:datatset}
\begin{tabular}{c|c|c|c}
\hline
                   & Recording Domain & Recording Environment & SNR  \\\hline
LibriTTS~\cite{LibriTTS} & AudioBook  & Multiple Rooms &   $6 \sim 30dB$ \\
VCTK~\cite{VCTK} & Reading  &  Hemi-anechoic Room  &  $15 \sim 37dB$  \\ 
CMU Arctic~\cite{CMU-Arctic} & Reading  & Sound Proof Room  & $18 \sim 45dB$ \\
Hi-Fi TTS~\cite{hifi} & AudioBook  & Multiple Rooms & $20 \sim 35dB$ \\ \hline

\end{tabular}

%\vspace{-15pt}
\end{table}

\begin{table}[ht]
\centering
%\scriptsize
\setlength{\tabcolsep}{0.4mm}
\renewcommand\arraystretch{2}
% \captionsetup[table]{labelfont={color=blue}, textfont={color=blue}}
\caption{The differences of the speaker modeling methods used by comparison systems.}
% \caption{Validation of each level TCR block.}
%\vspace{-5pt}
\label{tab:commodel}
\begin{tabular}{c|c|c|c|c|c}
\hline              &\makecell[c]{\textit{U}tterance- \\or \textit{M}ulti-level}  & SV-based  & Time-varying  & Channel-aware &Learnable \\\hline
MediumVC &  \textit{U}  & \checkmark  & $\times$  & $\times$ & $\times$ \\
 SRDVC &  \textit{U} &  $\times$  &  $\times$  & $\times$ & \checkmark\\ 
FragmentVC & \textit{M}  & $\times$ & \checkmark  & $\times$ &\checkmark \\
MTCR-VC & \textit{M}  & \checkmark & \checkmark & \checkmark & \checkmark \\ \hline

\end{tabular}

%\vspace{-15pt}
\end{table}
% \captionsetup[table]{labelfont={color=blue}, textfont={color=black}}

\subsection{Implementation Details}

All speech utterances are re-sampled to 24KHz, and an 80-dim mel spectrogram is computed with a 50ms frame length and 10ms frameshift. The ASR encoder output BNF has a dimension of 256. We use Pyworld\footnote{https://github.com/JeremyCCHsu/Python-Wrapper-for-World-Vocoder} to extract 1-dim logarithmic-domain fundamental frequency (\textit{lf0}). The lf0 normalized by utterance-level min-max method is used as the pitch contour input of the pitch encoder. And the down-sampling rate in the pitch encoder is 8. The x-vector from the SV model is a 192-dim feature. In the MTCR speaker module, the scale factors $\gamma_t$ and $\gamma_c$ are both set to 4, and the temporal ranges $\gamma_{tr}$ from low level (top TCR block in Fig.~\ref{fig:MTCR-VC}) to high level (bottom TCR block in Fig.~\ref{fig:MTCR-VC}) are 16, 4, 1, which keep the temporal range at a length of 640ms. During training, we set the loss weights as $\lambda_{mel}=4$, $\lambda_{sty}=0.1$, $\lambda_{con}=0.01$, $\lambda_{spk}=0.1$, based on empirical experience. We train MTCR-VC for 200 epochs with a batch size of 128. The learning rate is set to 1e-5 and decays every 50,000 steps with a decay rate of 0.5. The prenet in the MTCR speaker module consists of a single convolution layer. The configuration of convolution layers in the MTCR speaker module and fusion blocks of the speech decoder is 3$\times$3 filter with 1$\times$1 stride and replication padding. The speech decoder keeps the same settings for the fusion blocks, smoother, and postnet as \cite{FragmentvcAVLin2021}. For perceptual constraints, following \cite{wang2022delivering}, the style model comprises a reference encoder and 3 FC layers, while the content model adopts a CBHG structure~\cite{TacotronWang2017}.

\begin{table*}[h]
\centering
% %\scriptsize
\setlength{\tabcolsep}{2.5mm}
\renewcommand\arraystretch{2}
%\vspace{-5pt}
\caption{Results of subjective and objective evaluations for zero-shot VC on ID and OOD speakers.}
% \caption{Comparison of different models in terms of subjective and objective evaluations.}
%\vspace{-5pt}
\label{exp:mos&obj}
\begin{tabular}{c|c|cc|cc|cc|cc|cc}
\hline
            \multirow{2}{*}{Method} &\multirow{2}{*}{Size (M)} & \multicolumn{2}{c|}{NMOS $(\uparrow)$}  & \multicolumn{2}{c|}{SMOS $(\uparrow)$} & \multicolumn{2}{c|}{WER $(\downarrow)$}  & \multicolumn{2}{c|}{\makecell[c]{$P_{lf0}$ $(\uparrow)$}}  &   \multicolumn{2}{c}{ACC $(\uparrow)$}      \\ \cline{3-12}
 &      & \multicolumn{1}{c|}{ID} & OOD & \multicolumn{1}{c|}{ID} & OOD & \multicolumn{1}{c|}{ID} & OOD & \multicolumn{1}{c|}{ID} & OOD & \multicolumn{1}{c|}{ID} & OOD  \\ \hline
MediumVC & 32.1  & \multicolumn{1}{l|}{\textbf{3.73$\pm$0.08}} & \textbf{3.65$\pm$0.07} & \multicolumn{1}{l|}{3.58$\pm$0.06} & 3.40$\pm$0.07 & \multicolumn{1}{l|}{2.187} & 2.041 & \multicolumn{1}{l|}{0.669} & 0.638  & \multicolumn{1}{l|}{0.949} & 0.903 \\
SRDVC   &  31.8   & \multicolumn{1}{l|}{3.54$\pm$0.08} & 3.57$\pm$0.08 & \multicolumn{1}{l|}{3.20$\pm$0.10} & 3.13$\pm$0.07  & \multicolumn{1}{l|}{1.848} & 1.921 & \multicolumn{1}{l|}{0.673} & 0.677  & \multicolumn{1}{l|}{0.875} & 0.742 \\
FragmentVC & 47.8    & \multicolumn{1}{l|}{3.47$\pm$0.07} & 3.50$\pm$0.08 & \multicolumn{1}{l|}{3.42$\pm$0.09} & 3.31$\pm$0.08 & \multicolumn{1}{l|}{1.926} & 2.103 & \multicolumn{1}{l|}{0.613} & 0.568  & \multicolumn{1}{l|}{0.927} & 0.865 \\
MTCR-VC & 23.4 & \multicolumn{1}{l|}{3.61$\pm$0.06} & 3.59$\pm$0.07 & \multicolumn{1}{l|}{\textbf{3.70$\pm$0.08}} & \textbf{3.68$\pm$0.07} & \multicolumn{1}{l|}{\textbf{1.745}} & \textbf{1.853} & \multicolumn{1}{l|}{\textbf{0.680}} & \textbf{0.701}  & \multicolumn{1}{l|}{\textbf{0.963}} & \textbf{0.958}  \\ \hline
\end{tabular}
%\vspace{-10pt}
\end{table*}

\subsection{Comparison Models}

% To evaluate the performance of MTCR-VC on the zero-shot VC task, we compare it with three recent representative zero-shot VC models with different speaker modeling methods. Notably, the same vocoder is utilized to reconstruct waveform from the Mel spectrogram for all the VC models. Details of these comparison models are introduced as follows.

To evaluate the performance of the proposed MTCR-VC
on the zero-shot VC task, we compare it with three recent representative zero-shot VC models with different speaker modeling methods, including speaker embedding extracted from the
SV, utterance-level speaker representation learning, and multi-
level speaker representation learning. The differences of speaker modeling methods used by comparison models are listed in Table~\ref{tab:commodel}. Notably, all VC models are trained on the same dataset, and the same vocoder is used to reconstruct waveform from the mel spectrogram for all the VC models. Details of these comparison models are introduced as follows.

\textbf{MediumVC}~\cite{mediumvc} adopts speaker representation extracted from a pre-trained SV model. It divides zero-shot VC (any-to-any) into any-to-one and one-to-any VC by a two-stage strategy. Subsequently, the any-to-one VC first converts the source speech to the speech of a specific non-target speaker, and then one-to-any VC converts the resulting speech to the target speech conditioned on the speaker representation of the target speaker's speech. We build both the two-stage VC models using a content encoder and a speech decoder, which have structures similar to those of MTCR-VC. However, the cross-attention in the fusion block is removed, and the speaker representation is added in each fusion block.

\textbf{SRDVC}~\cite{SRDVC} models speaker timbre by learning a utterance-level speaker representation. This model utilizes MI and classification losses to decompose speech into linguistic content, speaking style, and utterance-level speaker timbre. We use the officially released open-source code\footnote{https://github.com/YoungSeng/SRD-VC} to implement SRDVC.

\textbf{FragmentVC}~\cite{FragmentvcAVLin2021} learns variable-length speaker representations from multiple layers and feeds them to the corresponding decoder layers. We use the officially released open-source code\footnote{https://github.com/yistLin/FragmentVC} to implement FragmentVC. To ensure a fair comparison, we use BNF as the input instead of the wav2vec feature to avoid any potential speaker timbre leakage from the source speech.

% Three recent SOTA systems, which represent three typical zero-shot VC approaches, are compared with our MTCR-VC in the experiments. (1)~\textbf{MediumVC}~\cite{mediumvc}, which uses SV's speaker embedding to achieve zero-shot VC. (2)~\textbf{SRDVC}~\cite{SRDVC}, which models speaker timbre at utterance level and decomposes speech into different representations with MI loss and classification-based guidance. (3)~\textbf{FragmentVC}~\cite{FragmentvcAVLin2021}, which models speaker timbre at fine-grained level and generates speech with content-based fusion. For a fair comparison, the SV model used in MediumVC is the same as ours. SRDVC\footnote{https://github.com/YoungSeng/SRD-VC} and FragmentVC\footnote{https://github.com/yistLin/FragmentVC} are implemented using the officially released open-source code. Instead of wav2vec, BN is used as the input of FragmentVC to avoid speaker timbre leakage from source speech. All systems use the same vocoder to reconstruct waveform. 

%\vspace{-5pt}
\subsection{Evaluation Metrics}
% %\vspace{-5pt}
% To evaluate the performance of VC models in zero-shot task, we conduct both objective and subjective evaluation methods in the experiments.

\textbf{Subjective Metrics.} Following the typical mean opinion score (MOS) criterion, listeners rate the given speech with a score ranging from 1 to 5 for its speaker similarity (\textbf{SMOS}) or speech naturalness (\textbf{NMOS}). A higher score means better performance. A score value of 1 means very bad, and 5 means excellent. In the experiments, we randomly select 120 utterances from the testing set for subjective evaluations. A group of 20 listeners participates in the subjective listening MOS test. MOS is calculated with $95\%$ confidence intervals.
% ($10*12=120$),

\textbf{Objective Metrics.} For objective evaluations, we calculate word error rate (\textbf{WER}) via the ASR model to validate the intelligibility of the converted speech. Moreover, we calculate the Pearson correlation coefficient of \textit{lf0} between the source speech and converted speech to measure the consistency of speaking style, denoting $\boldsymbol{P_{lf0}}$. As in previous work~\cite{FragmentvcAVLin2021}, speaker accuracy (\textbf{ACC}) is calculated using an SV model to measure whether the converted speech is from the target speaker. The accuracy is determined by the cosine similarity between the SV's embeddings of two speech utterances exceeding a predefined threshold, based on the SV model's equal error rate (EER) over the considered dataset. Better speaker similarity would result in higher speaker accuracy.

% Better speaker similarity would result in higher SV accuracy.

\section{Experimental Results}
\label{sc:results}

% In this section, we present the results of subjective and objective evaluations and the ablation study of the proposed model. Further analysis of the MTCR speaker module is also discussed.

This section presents the results of subjective and objective evaluations and the ablation study conducted on the proposed model. Additionally, a detailed analysis of the MTCR speaker module is provided.

\subsection{Subjective and Objective Evaluations}

\subsubsection{Subjective Evaluation}
Table~\ref{exp:mos&obj} presents the NMOS and SMOS results for zero-shot VC on ID and OOD speakers. Regarding speech naturalness, our MTCR-VC outperforms SRDVC and FragmentVC, both of which model speaker timbre directly from speech during training. Meanwhile, Medium-VC gets better naturalness than MTCR-VC, likely due to its use of utterance-level speaker modeling and noise-robust speaker verification (SV) model. Specifically, directly using the x-vector extracted from the noise-robust SV model as the speaker representation results in less noisy converted speech and better speech naturalness.

Regarding speaker similarity, SMOS is degraded from ID to OOD speakers for all models. This degradation suggests that the performance is affected due to a mismatch between the training and testing sets. However, MTCR-VC gets better speaker similarity and a smaller SMOS gap (0.02) between ID and OOD speakers, effectively mitigating the mismatch and achieving robust performance in zero-shot VC. Additionally, it can be found that FragmentVC with fine-grained modeling outperforms SRDVC, demonstrating that the coarse utterance-level speaker modeling is insufficient for zero-shot VC. However, without considering the varying speaker information along the temporal and frequency dimensions in the speaker modeling process, FragmentVC is difficult to perform well. On the other hand, MediumVC achieves relatively better speaker similarity than the other two comparison models but suffers from the insufficient representation of the x-vector for the perceptual speaker timbre in the sense of human hearing. These findings highlight the superior performance of MTCR-VC for speaker modeling while preserving good speech naturalness.
% Besides, it can found that SRDVC gets poor results, compared with FragmentVC and MTCR-VC. It demonstrates that the modeling speaker

% 2.287  2.153    0.684 0.653   0.853 0.801

\begin{table}[h]
\centering
% %\scriptsize
\setlength{\tabcolsep}{2.5mm}
\renewcommand\arraystretch{2}
%\vspace{-8pt}
\caption{Results of ablation studies.}
%\vspace{-5pt}
% \caption{Ablation analysis of MTCR module and the cycle-based training strategy.}
\label{exp:ablation}
\begin{tabular}{c|cc|cc|cc}
\hline
           \multirow{2}{*}{Method}  & \multicolumn{2}{c|}{WER $(\downarrow)$}  & \multicolumn{2}{c|}{\makecell[c]{$P_{lf0}$ $(\uparrow)$}}  &   \multicolumn{2}{c}{ACC $(\uparrow)$} \\ \cline{2-7}
      & \multicolumn{1}{c|}{ID} & OOD & \multicolumn{1}{c|}{ID} & OOD & \multicolumn{1}{c|}{ID} & OOD   \\ \hline
MTCR-VC  & \multicolumn{1}{l|}{1.745} & 1.853 & \multicolumn{1}{l|}{0.680} & 0.701  & \multicolumn{1}{l|}{0.963} & 0.958 \\ \hline
\multicolumn{7}{c}{w/o MTCR Speaker Moudle} \\ \hline
\makecell[c]{w/o MTCR\\(Conv)}   & \multicolumn{1}{l|}{2.197} & 2.061 & \multicolumn{1}{l|}{0.617} & 0.609  & \multicolumn{1}{l|}{0.817} & 0.884 \\
\makecell[c]{w/o MTCR\\(SV)}  & \multicolumn{1}{l|}{2.287} & 2.153 & \multicolumn{1}{l|}{0.684} & 0.653  & \multicolumn{1}{l|}{0.853} & 0.801 \\
\hline
\multicolumn{7}{c}{w/o Cycle-based Training Strategy} \\ \hline
w/o \textit{Cycle} & \multicolumn{1}{l|}{1.932} & 1.869 & \multicolumn{1}{l|}{0.688} & 0.677  & \multicolumn{1}{l|}{0.956} & 0.903 \\
\hline
\multicolumn{7}{c}{w/o Perceptual Constraints} \\ \hline
w/o $L_{sty}$       & \multicolumn{1}{l|}{1.919} & 1.916 & \multicolumn{1}{l|}{0.681} & 0.691  & \multicolumn{1}{l|}{0.956} & 0.884 \\
w/o $L_{con}$     & \multicolumn{1}{l|}{2.057} & 2.145 & \multicolumn{1}{l|}{0.704} & 0.729  & \multicolumn{1}{l|}{0.909} & 0.889 \\
w/o $L_{spk}$ & \multicolumn{1}{l|}{1.937} & 1.906 & \multicolumn{1}{l|}{0.693} & 0.704  & \multicolumn{1}{l|}{0.935} & 0.927 \\  \hline
\end{tabular}
%\vspace{-10pt}
\end{table}

\subsubsection{Objective Evaluation}

Table~\ref{exp:mos&obj} presents the objective results in the last three columns. As illustrated, similar to the subjective results, the SMOS gap between ID and OOD speakers is also reflected in the objective metric of speaker accuracy. Notably, our proposed MTCR-VC achieves the highest speaker accuracy for both ID and OOD speakers. Additionally, MTCR-VC obtains the best performance on WER and $P_{lf0}$, indicating better speech intelligibility and speaking style consistency. Consistent with subjective results, the objective results also demonstrate that MTCR-VC outperforms other models in speaker similarity while preserving good speech naturalness.

% Among the four models, SRDVC and MTCR-VC, which decompose speech into linguistic content, speaking style, and speaker timbre, get better WER and $P_{lf0}$ than the others.% Moreover, our proposed method MTCR-VC obtains the best performance on all three objective metrics, demonstrating that MTCR-VC is effective for zero-shot VC and gets better speech intelligibility, speaking style consistency and speaker accuracy. Subjective and objective experimental results show that MTCR-VC achieves new SOTA performance in speaker similarity while preserving good speech naturalness.
% The objective comparison result of different models on speaker accuracy also presents similar results to that obtained in the subjective evaluation. The performance gap between ID and OOD speakers is also reflected in the objective result.

% \subsection{Effectiveness Verification}

% Above, experimental results

\subsection{Ablation Study}

As shown in Table~\ref{exp:ablation}, We conduct further studies to assess the effectiveness of MTCR-VC with ablations on the MTCR speaker module, cycle-based training strategy, and perceptual constraints.

% Specifically, we replace the MTCR speaker module with three convolution layers, as in Lin et al.~\cite{FragmentvcAVLin2021}, to evaluate its efficacy, forming the model \textit{w/o MTCR}. Instead of the cycle-based training strategy, we remove the unpaired path in the model (\textit{w/o Cycle}), and different speech utterances from the same speaker are selected to perform the training process as in previous work~\cite{FragmentvcAVLin2021,s2vc}. Furthermore, we build the models \textit{w/o $L_{sty}$},~\textit{w/o $L_{con}$}, and \textit{w/o $L_{spk}$} by removing the perceptual constraints of style, content, and speaker, respectively, to verify their effectiveness.

\subsubsection{MTCR Speaker Module}

% 为了验证模型的有效性
To confirm the efficacy of the MTCR speaker module, we replace it with three convolution layers, as in Lin et al.~\cite{FragmentvcAVLin2021}, forming the model \textit{w/o MTCR~(Conv)}. Besides, we also implement a model \textit{w/o MTCR~(SV)}, which only uses SV embedding to represent speaker timbre to involve this ablation. As can be seen in Table~\ref{exp:ablation}, discarding the MTCR speaker module leads to a noticeable decrease in speaker accuracy. Furthermore, the absence of the MTCR module weakens the disentanglement ability, resulting in higher WER and lower $P_{lf0}$ scores. These findings indicate the effectiveness of the MTCR speaker module for speaker modeling and its positive impact on zero-shot VC.

\subsubsection{Cycle-based Training Strategy}
Instead of the cycle-based training strategy, we implement an ablation model (\textit{w/o Cycle}) using a popular training strategy in previous work~\cite{FragmentvcAVLin2021,s2vc}, in which different speech utterances from the same speaker are selected to perform the training process. We can observe that replacing the cycle-based training strategy leads to performance decreases in WER, $P_{lf0}$, and ACC. This demonstrates the cycle-based training strategy employed in this work is helpful for zero-shot VC.

\subsubsection{Perceptual Constraints}
Table~\ref{exp:ablation} shows the effect of perceptual losses we used on MTCR-VC. As can be seen, the constraints of style, content, and speaker play important roles. Dropping these constraints brings corresponding performance degradation in $P_{lf0}$, WER, and ACC, respectively. The result indicates that the three perceptual constraints effectively improve speech disentanglement performance in the zero-shot VC task. Among them, \textit{w/o $L_{spk}$} cause small degradation to speaker accuracy because the cycle-based training process implicitly ensures the consistency of speaker timbre.

% \textcolor{red}{add expermen result}
% The results of ablation studies are presented in Table~\ref{exp:ablation}. As can be seen, discarding the MTCR speaker module leads to a noticeable decrease in speaker accuracy. Furthermore, the absence of the MTCR module weakens the disentanglement ability of \textit{w/o MTCR}, resulting in higher WER and lower $P_{lf0}$ scores. These findings indicate the effectiveness of the MTCR speaker module for speaker modeling and its positive impact on zero-shot VC. Replacing the cycle-based training strategy leads to performance decreases in WER, $P_{lf0}$, and ACC. This demonstrates the cycle-based training strategy employed in this work is helpful for zero-shot VC. Moreover, dropping the perceptual constraints of style, content, and speaker brings corresponding performance degradation in $P_{lf0}$, WER, and ACC, respectively. The result indicates that the three perceptual constraints effectively improve speech disentanglement performance in the zero-shot VC task. Among them, \textit{w/o $L_{spk}$} cause small degradation to speaker accuracy because the cycle-based training process implicitly ensures the consistency of speaker timbre.

% the disentanglement of style, linguistic content, and speaker timbre. 

\begin{figure}[ht]	
% %\vspace{-5pt}
\centering
% %\vspace{-5pt}
\includegraphics[width=1.0\linewidth]{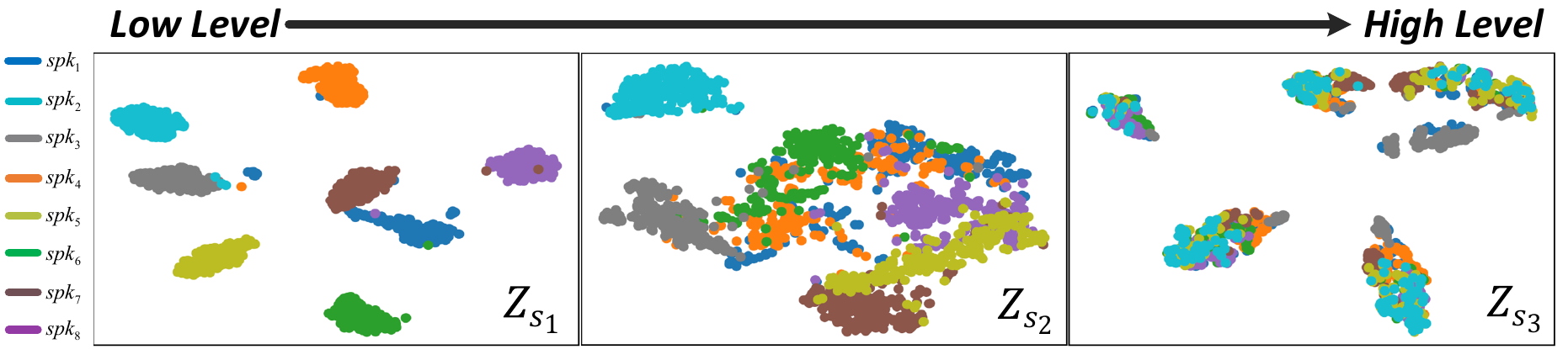}
% %\vspace{-5pt}
\caption{Visualization of speaker representations in each level. Different colors represent different speakers.}
%\vspace{-17pt}
\label{fig:tsne}
\end{figure}
\begin{table}[ht]
\centering
% %\scriptsize
\setlength{\tabcolsep}{2.5mm}
\renewcommand\arraystretch{2}
\caption{Validation results of multi-level TCR blocks.}
% \caption{Validation of each level TCR block.}
%\vspace{-5pt}
\label{exp:level}
\begin{tabular}{c|c|c|c}
\hline
                   & w/o $Z_{s_{3}}$ & w/o $Z_{s_{2}}$\&$Z_{s_{3}}$ & MTCR-VC \\\hline
\makecell[c]{WER $(\downarrow)$}   & 1.98 &   2.258 & 1.648 \\
\makecell[c]{ACC $(\uparrow)$}   & 0.837  &  0.657  & 0.915 \\ \hline
\end{tabular}
%\vspace{-15pt}
\end{table}

\subsection{Analysis of the MTCR Speaker Module}
To further investigate the effectiveness of the MTCR speaker module, we analyze the speaker representation learned in each level of the TCR blocks and explore the speaker retrieval process in each TCR block. \textit{Visualization} and \textit{quantitative metrics} both are conducted to analyze these two aspects.
% In this section, we further explore the behavior of the MTCR speaker module in our model. Layer-wise analysis will be conducted to speaker representation and retrieval process. 

%\subsubsection{Speaker Representation} 
\subsubsection{Speaker Representation} 
To explore what is learned in the speaker representation, we first visualize the speaker representations extracted from each TCR block with t-SNE. Since the speaker representation is of variable lengths and this visualization can only be done on a single vector, the speaker representation is averaged along the temporal axis to obtain a single vector and then visualized from a \textit{global perspective}. We select 8 speakers from the test set for this visualization, each contributing 200 utterances. As shown in Fig.~\ref{fig:tsne}, speaker representations from the low-level TCR block (top TCR block in Fig.~\ref{fig:MTCR-VC}) group into 8 speaker clusters, indicating the presence of explicit speaker discriminative information. As the layer deepens, the speaker clusters start to overlap, eventually becoming indistinguishable. The low-level block is close to the speech spectrogram and thus can easily capture more speaker timbre information. In contrast, the high-level block captures much finer speaker-related characteristics but less speaker-discriminative information. This phenomenon is consistent with a previous zero-shot study~\cite{unetts}. Please note that this visualization only analyzes the representation from \textit{global perspective}, but the speaker representations used in the MTCR-VC are \textit{time-varying} features.

Therefore, to further investigate the impact of different TCR blocks, we objectively evaluate the performance of MTCR-VC by sequentially discarding blocks from the high-level layer to the low-level layer. Specifically, the variant model \textit{w/o $Z_{s_3}$} only uses the first two TCR blocks in the speaker module and is trained from scratch to verify the role of $Z_{s_3}$, compared to the proposed MTCR-VC. Besides, discarding both second and third TCR blocks is also verified, forming the model \textit{w/o $Z_{s_2}$ \& $Z_{s_3}$}. The validation results are presented in Table~\ref{exp:level}. As can be seen, discarding the use of the third block in the model \textit{w/o $Z_{s_3}$} leads to decreased speaker accuracy. It suggests that the third block also learns time-varying speaker representation, which contains fine-grained speaker information, even though the averaged $Z_{s_3}$ shows little discrimination among different speakers from the global perspective in Fig~\ref{fig:tsne}. Furthermore, removing the second block and third block in the model \textit{w/o $Z_{s_2}$ \& $Z_{s_3}$} leads to more performance degradation in speaker accuracy. This indicates that the speaker representations learned in $Z_{s_2}$ and $Z_{s_3}$ are important for speaker accuracy in MTCR-VC. Additionally, it can be found that discarding the use of TCR blocks increases the WER. This outcome occurs because the speaker representation participates in the speech representation fusion, and the performance of speaker modeling affects the speech generation process.

% one block's output $Z_{s_l}$ with zero embedding and keeping the other blocks' outputs. The validation results of each TCR block are presented in Table~\ref{exp:level}. Notably, the results for the first block (w/o $Z_{s_1}$) are not listed in the table since dropping the output of the first block leads to abnormal noise instead of producing comprehensible speech. It suggests that the speaker representation learned in $Z_{s_1}$ is an essential acoustic feature for speech generation. Furthermore, removing the speaker representations learned in the second block and third block ($Z_{s_2}$ and $Z_{s_3}$) results in decreased speaker accuracy. This indicates that the speaker representations learned in $Z_{s_2}$ and $Z_{s_3}$ contain fine-grained speaker information that is important for speaker accuracy. Additionally, it can be found that dropping the learned speaker representation increases the WER. This outcome occurs because the speaker representation participates in the speech representation fusion, and the dropping affects the speech generation process.

% Since the content-specific fusion process is affected by the removal, WER gets higher. These results show that the speaker representation learned in different levels contains different characteristic information of speaker timbre.
%information in different aspects.

\begin{figure*}[ht]
\centering
\begin{minipage}{0.3\linewidth}
    \subfigure[Spectrogram of a speaker utterance]{
      \includegraphics[width=1\columnwidth]{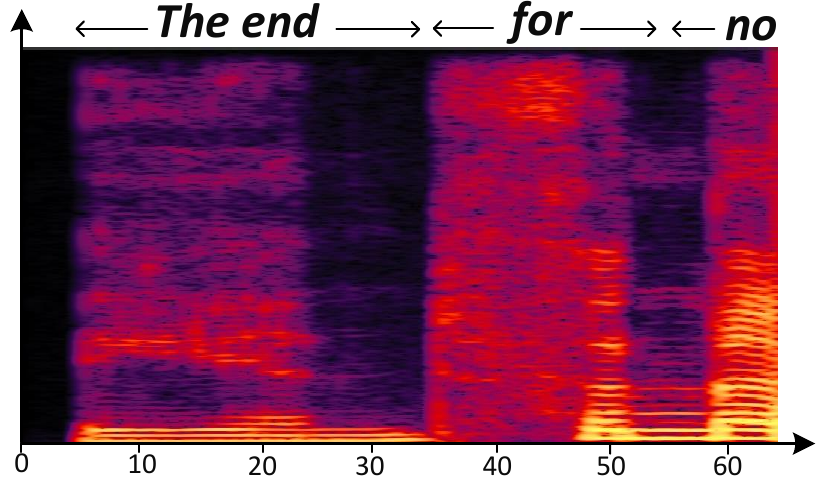}}
      \\
      \centering
    \subfigure[Pitch contour of the utterance]{
      \includegraphics[width=0.8\columnwidth]{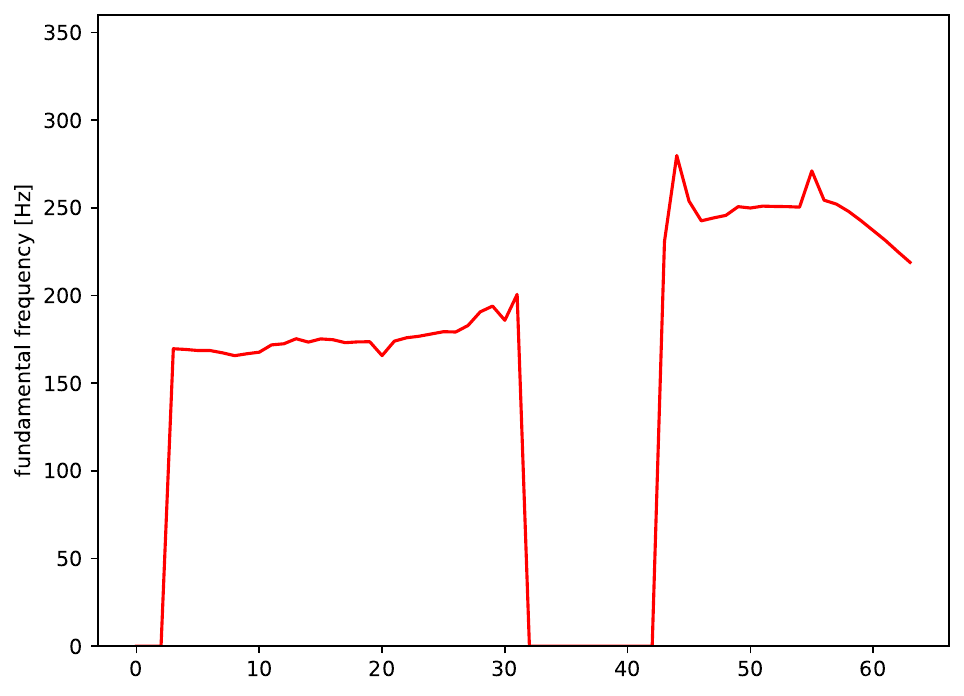}}
\end{minipage}
\begin{minipage}{0.65\linewidth}
    \subfigure[Channel retrieval]{
      \includegraphics[width=1\columnwidth]{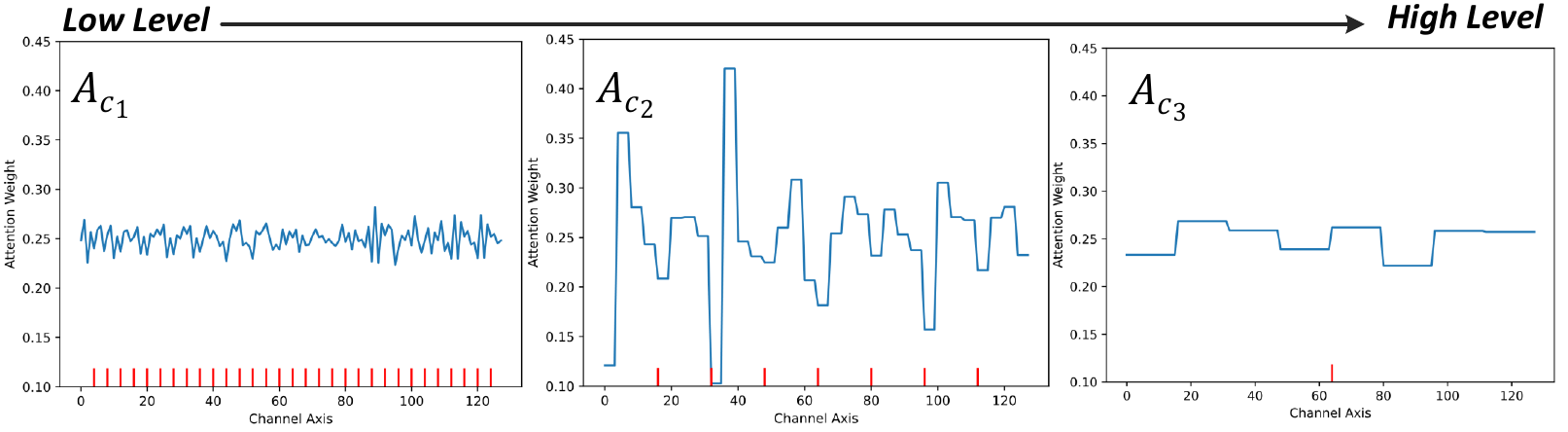}}\\
    \subfigure[Temporal retrieval]{
      \includegraphics[width=1\columnwidth]{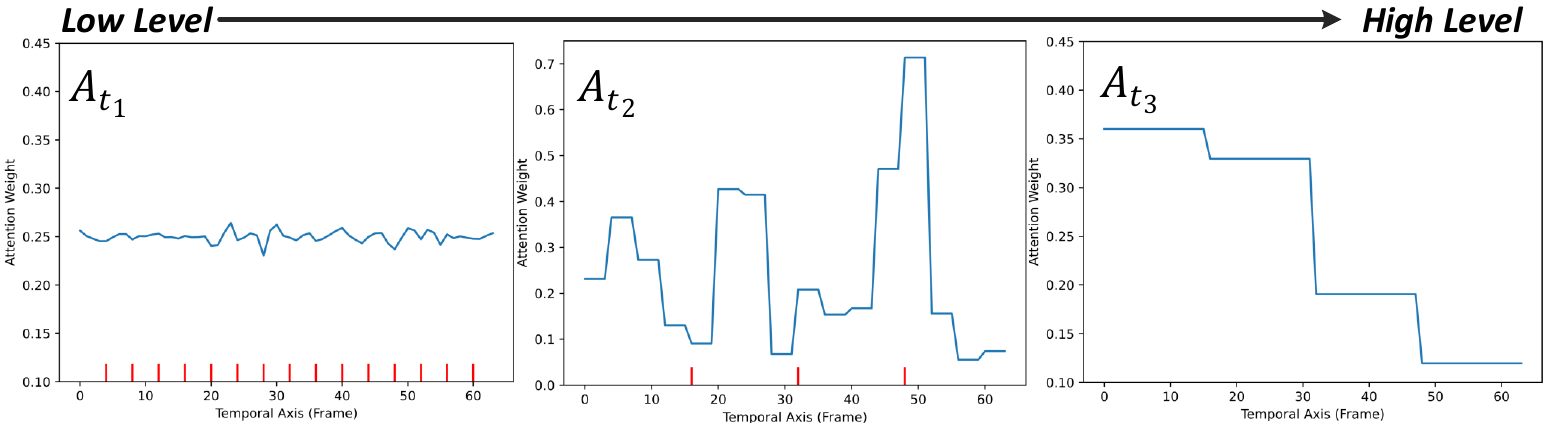}}
\end{minipage}

\caption{Visualization of attention map in each level retrieval. (a) Speaker speech. (b) Pitch contour. (c) Channel retrieval. (d) Temporal retrieval. Note that the speaker's speech with 64 frames is part of a long utterance, and the corresponding content in the speech is \textit{``The end for no"}. The intervals between red lines represent different segments. And in (c) and (d), arrows from left to right indicate the levels from low to high.}
\label{fig:attn}
\end{figure*}

%\subsubsection{Speaker Retrieval Process} 
\subsubsection{Speaker Retrieval Process} 
As described in Section~\ref{sec:retri}, retrieving speaker information in temporal and channel dimensions with multiple granularities is crucial for speaker modeling in zero-shot voice conversion. Therefore, we further investigate the speaker retrieval process in temporal and channel dimensions at multi-level TCR blocks. The attention map is a vital component in the retrieval process. Hence, we first plot the attention map $A_{t_{l}}$ and $A_{c_{l}}$ learned in temporal and channel retrieval from low-level TCR block to high-level TCR block. Specifically, we use one speech utterance as the input of the MTCR speaker module to obtain the attention map. Fig.~\ref{fig:attn} shows the spectrogram of the speech utterance, corresponding pitch contour, and the attention map. In Fig.~\ref{fig:attn} (c) and (d), the red line in the horizontal axis marks the boundary of each segment. Note that the attention map of each temporal or channel segment has the dimension of $1 \times\gamma_{t} $ or $1 \times \gamma_{c}$. For visualization, we flat and repeat the $A_{t_{l}}$ and $A_{c_{l}}$ to the origin temporal and channel length, respectively. The vertical axis value ranges from 0 to 1 and represents the proportion of each feature in the temporal or channel dimensions within a segment. This proportion ensures that the sum of weights within each segment is 1. As can be seen, in the low-level block, the fluctuation of attention weight in both temporal and channel retrieval is relatively small since the adjacent temporal and channel features within a segment are similar. As the layer deepens, the attention weight value begins to change sharply, indicating that the model is more sensitive to the channel and temporal regions that are highly correlated to the speaker's SV embedding (e.g. speaker timbre information). For instance, in $A_{t_{2}}$, the attention weight increases when the speaker begins to speak in the fifth frame. In the third segment, from frame 32 to 48, compared with the unvoiced part, the voiced sound (from 44 to 48) with higher attention weight provides more speaker information for the model. In voice sound from frames 48 to 64, the attention weights change according to the correlation to the speaker timbre. These observations suggest that the speaker retrieval process enables the model to flexibly adjust its attention across temporal and channel regions, considering variations in the amount of speaker information present.

% The model pays attention to the channel regions that are highly correlated to the speaker timbre and focuses on the period when speaker information is richer in the temporal dimension. 
% via plotting and replacing the attention map learned in the process.

%  because it is vital in the retrieval process
% As the number of layers deepens, the value of the attention weight begins to change drastically

% 在低维度时，attion在时域和通道域的数值变化小。这是由于Since the high similarity between adjacent temporal features and channel features in low-level segment。随着层数的变深，attention权重的数值开始剧烈变化。在通道上，模型将权重放到不同的通道上去找出更加说话人相关的区域。时序上，模型关注什么时候说话人信息更丰富，比如在图(c)At2, 在 the weightcurve 0-16 人开始说话时取得了更高的权重  and 32-48表明模型认为相比浊音清音提供了更多的说话人信息 。最终，和At2heAc2相比，注意力权重变得稍微平缓。这表明这个过程帮助模型根据说话人信息的含量去灵活调整不同时序和通道特征的比例。
\begin{table}[h]
\centering
%\vspace{-5pt}
%\scriptsize
\caption{Validation results of the speaker retrieval process.}
% \caption{Validation of each level retrieval process.}
%\vspace{-5pt}
\label{exp:att}
\setlength{\tabcolsep}{1mm}
\renewcommand\arraystretch{2}
\begin{tabular}{c|c|c|c|c|c|c}
\hline
                   & Block 1 & Block 2 & Block 3 
 & Temporal & Channel &MTCR-VC\\\hline
\makecell[c]{WER $(\downarrow)$} & 1.722  & 1.801  &  1.882 & 2.021  & 1.836 &1.648 \\
\makecell[c]{ACC $(\uparrow)$} & 0.842  & 0.783 &  0.817 & 0.744 & 0.837 &0.915\\ \hline
\end{tabular}
%\vspace{-10pt}
\end{table}

% b1 1.5 1.65 1.43 1.90   1.62
% b2 1.71 2.09 1.87 1.53 1.8
% b3 1.81 1.78 1.87 2.46 1.98
% channel 1.78 2.09 1.90 2.31 2.02
% temp 1.5 1.87 1.62 1.93 1.73

% wb2b3 2.50 2.06 2.46 2.00 2.25
% wb3 1.65 1.96 2.18 1.81 1,9

To further quantify the effect of the speaker retrieval process, we average the attention map in a specific retrieval process while keeping the other retrieval unchanged. For example, to verify the retrieval process in the first TCR block, only the attention weights of the first block are set to the fixed average value, and the whole modified model is trained from scratch. As shown in Table.~\ref{exp:att}, we implement five variants to verify the efficiency of the retrieval process. Experiments are conducted on 8 unseen speakers mentioned above, and the validation results of the speaker retrieval process are shown in the table. The results show that discarding the retrieval process in any one of the TCR blocks leads to a degradation of performance on WER and ACC. Compared with the retrieval process in block 2 and block 3, averaging the attention map in block 2 shows a relatively obvious degradation in speaker accuracy, which indicates that the retrieval process in block 2 is important for speaker modeling. Moreover, as shown in the fifth and sixth columns of Table~\ref{exp:att}, averaging attention maps in temporal retrieval or channel retrieval in all blocks apparently shows an increase in WER and a decrease in speaker accuracy. This result suggests that temporal and channel retrieval help capture speaker timbre. Interestingly, changing temporal retrieval has a relatively large degradation in intelligibility because the attention map $A_{t_{l}}$ learned in the temporal retrieval is used for content-based alignment in the fusion block, as shown in Fig.~\ref{fig:MTCR-VC}.

\section{Conclusion}
\label{sc:conclusion}
This paper introduces MTCR-VC, a new zero-shot VC model with a novel speaker modeling approach. Specifically, we propose the MTCR speaker module, which hierarchically queries speaker timbre from the target speaker's speech across different temporal and channel granularities by stacking multiple TCR blocks. This allows the model for flexible allocation of attention across temporal and channel regions based on the richness of speaker information at different stages, thus improving the robustness of speaker timbre modeling. Moreover, applying the cycle-based training strategy facilitates the process of speech reconstruction and
disentanglement toward the aim of the zero-shot VC task. Our experimental results show that the proposed MTCR-VC performs better in modeling speaker timbre while maintaining good speech naturalness.

Although the experimental results have demonstrated good performance on speaker timbre modeling, we have to point out that there are still some limitations. Specifically, the zero-shot VC performance, including naturalness and speaker similarity, varies among different speakers and speech recordings with different recording conditions, likely due to the limited speaker diversity and recording environments in the training corpus. To address this limitation, future work should consider modeling speaker timbre from more diverse speech data, containing even a larger number of speakers and speech recordings, to improve the generalization ability of zero-shot VC.
% Moreover, due to the application of content-based alignment in the speech representation fusion, the content coverage from the target speaker's speech and source speech may affect the zero-shot performance. Poor coverage may result in noise and unnatural pronunciation in the converted speech.


\bibliographystyle{IEEEtran}
\bibliography{ref.bib}

\begin{thebibliography}{10}
\providecommand{\url}[1]{#1}
\csname url@samestyle\endcsname
\providecommand{\newblock}{\relax}
\providecommand{\bibinfo}[2]{#2}
\providecommand{\BIBentrySTDinterwordspacing}{\spaceskip=0pt\relax}
\providecommand{\BIBentryALTinterwordstretchfactor}{4}
\providecommand{\BIBentryALTinterwordspacing}{\spaceskip=\fontdimen2\font plus
\BIBentryALTinterwordstretchfactor\fontdimen3\font minus \fontdimen4\font\relax}
\providecommand{\BIBforeignlanguage}[2]{{%
\expandafter\ifx\csname l@#1\endcsname\relax
\typeout{** WARNING: IEEEtran.bst: No hyphenation pattern has been}%
\typeout{** loaded for the language `#1'. Using the pattern for}%
\typeout{** the default language instead.}%
\else
\language=\csname l@#1\endcsname
\fi
#2}}
\providecommand{\BIBdecl}{\relax}
\BIBdecl

\bibitem{GANHsu2017VoiceCF}
C.-C. Hsu, H.-T. Hwang, Y.-C. Wu, Y.~Tsao, and H.-M. Wang, ``Voice conversion from unaligned corpora using variational autoencoding wasserstein generative adversarial networks,'' in \emph{International Speech Communication Association (Interspeech)}, 2017, pp. 3364--3368.

\bibitem{VAEHsu2016Voicevae}
C.-C. Hsu, H.-T. Hwang, Y.-C. Wu, Y.~Tsao, and H.~Wang, ``Voice conversion from non-parallel corpora using variational auto-encoder,'' in \emph{Asia-Pacific Signal and Information Processing Association Annual Summit and Conference (APSIPA)}, 2016, pp. 1--6.

\bibitem{PPGSun2016PhoneticPF}
L.~Sun, K.~Li, H.~Wang, S.~Kang, and H.~Meng, ``Phonetic posteriorgrams for many-to-one voice conversion without parallel data training,'' in \emph{International Conference on Multimedia and Expo (ICME)}, 2016, pp. 1--6.

\bibitem{maskcyclegan}
T.~Kaneko, H.~Kameoka, K.~Tanaka, and N.~Hojo, ``Maskcyclegan-vc: Learning non-parallel voice conversion with filling in frames,'' in \emph{International Conference on Acoustics, Speech and Signal Processing (ICASSP)}, 2021, pp. 5919--5923.

\bibitem{stargan}
H.~Kameoka, T.~Kaneko, K.~Tanaka, and N.~Hojo, ``Stargan-vc: Non-parallel many-to-many voice conversion using star generative adversarial networks,'' in \emph{Spoken Language Technology Workshop (SLT)}, 2018, pp. 266--273.

\bibitem{wang21g_interspeech}
Z.~Wang, X.~Zhou, F.~Yang, T.~Li, H.~Du, L.~Xie, W.~Gan, H.~Chen, and H.~Li, ``Enriching source style transfer in recognition-synthesis based non-parallel voice conversion,'' in \emph{International Speech Communication Association (Interspeech)}, 2021, pp. 831--835.

\bibitem{autovcqian2019autovc}
K.~Qian, Y.~Zhang, S.~Chang, X.~Yang, and M.~Hasegawa-Johnson, ``Autovc: Zero-shot voice style transfer with only autoencoder loss,'' in \emph{International Conference on Machine Learning (ICML)}, 2019, pp. 5210--5219.

\bibitem{speechsplit}
K.~Qian, Y.~Zhang, S.~Chang, M.~Hasegawa-Johnson, and D.~Cox, ``Unsupervised speech decomposition via triple information bottleneck,'' in \emph{International Conference on Machine Learning (ICML)}, 2020, pp. 7836--7846.

\bibitem{speakeraware}
Y.~Zhang, H.~Che, J.~Li, C.~Li, X.~Wang, and Z.~Wang, ``One-shot voice conversion based on speaker aware module,'' in \emph{IEEE International Conference on Acoustics, Speech and Signal Processing (ICASSP)}, 2021, pp. 5959--5963.

\bibitem{mediumvc}
Y.~Gu, Z.~Zhang, X.~Yi, and X.~Zhao, ``Mediumvc: Any-to-any voice conversion using synthetic specific-speaker speeches as intermedium features,'' \emph{Arxiv}, 2021.

\bibitem{INchou2019oneshot}
J.~chieh Chou and H.-Y. Lee, ``One-shot voice conversion by separating speaker and content representations with instance normalization,'' in \emph{International Speech Communication Association (Interspeech)}, 2019, pp. 664--668.

\bibitem{AgainVC}
Y.-H. Chen, D.-Y. Wu, T.-H. Wu, and H.-y. Lee, ``{Again-VC: A One-Shot Voice Conversion Using Activation Guidance and Adaptive Instance Normalization},'' in \emph{International Conference on Acoustics, Speech and Signal Processing (ICASSP)}, 2021, pp. 5954--5958.

\bibitem{SIGVC}
H.~Zhang, Z.~Cai, X.~Qin, and M.~Li, ``Sig-vc: A speaker information guided zero-shot voice conversion system for both human beings and machines,'' in \emph{Conference on Acoustics, Speech and Signal Processing (ICASSP)}, 2022, pp. 6567--65\,571.

\bibitem{avqvc}
H.~Tang, X.~Zhang, J.~Wang, N.~Cheng, and J.~Xiao, ``Avqvc: One-shot voice conversion by vector quantization with applying contrastive learning,'' in \emph{Conference on Acoustics, Speech and Signal Processing (ICASSP)}, 2022, pp. 4613--4617.

\bibitem{contrastive}
J.~Ebbers, M.~Kuhlmann, T.~Cord-Landwehr, and R.~Haeb-Umbach, ``Contrastive predictive coding supported factorized variational autoencoder for unsupervised learning of disentangled speech representations,'' in \emph{International Conference on Acoustics, Speech and Signal Processing (ICASSP)}, 2021, pp. 3860--3864.

\bibitem{nansy}
H.-S. Choi, J.~Lee, W.~Kim, J.~Lee, H.~Heo, and K.~Lee, ``Neural analysis and synthesis: Reconstructing speech from self-supervised representations,'' in \emph{Neural Information Processing Systems(NeurIPS)}, 2021, pp. 16\,251--16\,265.

\bibitem{VQMIVC}
D.~Wang, L.~Deng, Y.~T. Yeung, X.~Chen, X.~Liu, and H.~Meng, ``Vqmivc: Vector quantization and mutual information-based unsupervised speech representation disentanglement for one-shot voice conversion,'' in \emph{International Speech Communication Association (Interspeech)}, 2021, pp. 1344--1348.

\bibitem{SRDVC}
S.~Yang, M.~Tantrawenith, H.~Zhuang, Z.~Wu, A.~Sun, J.~Wang, N.~Cheng, H.~Tang, X.~Zhao, J.~Wang, and H.~Meng, ``Speech representation disentanglement with adversarial mutual information learning for one-shot voice conversion,'' in \emph{International Speech Communication Association (Interspeech)}, 2022, pp. 2553--2557.

\bibitem{MAP}
J.~Wang, J.~Li, X.~Zhao, Z.~Wu, S.~Kang, and H.~Meng, ``Adversarially learning disentangled speech representations for robust multi-factor voice conversion,'' \emph{Arxiv}, 2021.

\bibitem{CAVC}
R.~Xiao, X.~Xing, J.~Yang, and X.~Xu, ``Ca-vc: A novel zero-shot voice conversion method with channel attention,'' in \emph{Asia-Pacific Signal and Information Processing Association Annual Summit and Conference (APSIPA)}, 2021, pp. 800--807.

\bibitem{du21_interspeech}
H.~Du and L.~Xie, ``Improving robustness of one-shot voice conversion with deep discriminative speaker encoder,'' in \emph{International Speech Communication Association (Interspeech)}, 2021, pp. 1379--1383.

\bibitem{VQVC+}
D.-Y. Wu, Y.-H. Chen, and H.~yi~Lee, ``Vqvc+: One-shot voice conversion by vector quantization and u-net architecture,'' in \emph{International Speech Communication Association (Interspeech)}, 2020, pp. 4691--4695.

\bibitem{FragmentvcAVLin2021}
Y.~Y. Lin, C.~M. Chien, J.~hao Lin, H.~yi~Lee, and L.-S. Lee, ``Fragmentvc: Any-to-any voice conversion by end-to-end extracting and fusing fine-grained voice fragments with attention,'' in \emph{International Conference on Acoustics, Speech and Signal Processing (ICASSP)}, 2021, pp. 5939--5943.

\bibitem{retriever}
D.~Yin, X.~Ren, C.~Luo, Y.~Wang, Z.~Xiong, and W.~Zeng, ``Retriever: Learning content-style representation as a token-level bipartite graph,'' in \emph{International Conference on Learning Representations (ICLR)}, 2021.

\bibitem{attbasedzsl}
T.~Ishihara and D.~Saito, ``Attention-based speaker embeddings for one-shot voice conversion,'' in \emph{International Speech Communication Association (Interspeech)}, 2020, pp. 806--810.

\bibitem{Xuli_interspeech}
X.~Li, S.~Liu, and Y.~Shan, ``A hierarchical speaker representation framework for one-shot singing voice conversion,'' in \emph{International Speech Communication Association (Interspeech)}, 2022, pp. 4307--4311.

\bibitem{unet}
O.~Ronneberger, P.~Fischer, and T.~Brox, ``U-net: Convolutional networks for biomedical image segmentation,'' in \emph{Medical Image Computing and Computer-Assisted Intervention (MICCAI)}, 2015, pp. 234--241.

\bibitem{unetts}
R.~Li, D.~Pu, M.~Huang, and B.~Huang, ``Unet-tts: Improving unseen speaker and style transfer in one-shot voice cloning,'' in \emph{International Conference on Acoustics, Speech and Signal Processing (ICASSP)}, 2022, pp. 8327--8331.

\bibitem{audio}
B.~Gold, N.~Morgan, D.~Ellis, and D.~O'Shaughnessy, ``Speech and audio signal processing: Processing and perception of speech and music, second edition,'' \emph{The Journal of the Acoustical Society of America}, vol. 132, pp. 1861--2, 09 2012.

\bibitem{liu2022mfa}
T.~Liu, R.~K. Das, K.~A. Lee, and H.~Li, ``Mfa: Tdnn with multi-scale frequency-channel attention for text-independent speaker verification with short utterances,'' in \emph{International Conference on Acoustics, Speech and Signal Processing (ICASSP)}, 2022, pp. 7517--7521.

\bibitem{multi-frequency}
M.~Sang and J.~H. Hansen, ``Multi-frequency information enhanced channel attention module for speaker representation learning,'' in \emph{International Speech Communication Association (Interspeech)}, 2022, pp. 321--325.

\bibitem{ECAPA_TDNN}
B.~Desplanques, J.~Thienpondt, and K.~Demuynck, ``{ECAPA-TDNN: Emphasized Channel Attention, Propagation and Aggregation in TDNN Based Speaker Verification},'' in \emph{International Speech Communication Association (Interspeech)}, 2020, pp. 3830--3834.

\bibitem{human_voice}
Wikipedia, ``Human voice,'' \url{https://en.wikipedia.org/wiki/Human_voice}.

\bibitem{wang2022delivering}
Z.~Wang, X.~Wang, L.~Xie, Y.~Chen, Q.~Tian, and Y.~Wang, ``Delivering speaking style in low-resource voice conversion with multi-factor constraints,'' \emph{ArXiv}, 2022.

\bibitem{GMMStylianou1998}
Y.~{Stylianou}, O.~{Cappe}, and E.~{Moulines}, ``Continuous probabilistic transform for voice conversion,'' \emph{IEEE Transactions on Speech and Audio Processing}, vol.~6, no.~2, pp. 131--142, 1998.

\bibitem{GMMToda2007}
T.~{Toda}, A.~W. {Black}, and K.~{Tokuda}, ``Voice conversion based on maximum-likelihood estimation of spectral parameter trajectory,'' \emph{IEEE Transactions on Audio, Speech, and Language Processing}, vol.~15, no.~8, pp. 2222--2235, 2007.

\bibitem{NNDesai2009}
S.~{Desai}, E.~V. {Raghavendra}, B.~{Yegnanarayana}, A.~W. {Black}, and K.~{Prahallad}, ``Voice conversion using artificial neural networks,'' in \emph{International Conference on Acoustics, Speech and Signal Processing (ICASSP)}, 2009, pp. 3893--3896.

\bibitem{NNSun2015}
L.~{Sun}, S.~{Kang}, K.~{Li}, and H.~{Meng}, ``Voice conversion using deep bidirectional long short-term memory based recurrent neural networks,'' in \emph{International Conference on Acoustics, Speech and Signal Processing (ICASSP)}, 2015, pp. 4869--4873.

\bibitem{RAM}
V.~Mnih, N.~Heess, A.~Graves \emph{et~al.}, ``Recurrent models of visual attention,'' \emph{Neural Information Processing Systems(NeurIPS)}, vol.~27, 2014.

\bibitem{SEnet}
J.~Hu, L.~Shen, and G.~Sun, ``Squeeze-and-excitation networks,'' in \emph{Conference on computer vision and pattern recognition (CVPR)}, 2018, pp. 7132--7141.

\bibitem{TAM}
Z.~Liu, L.~Wang, W.~Wu, C.~Qian, and T.~Lu, ``Tam: Temporal adaptive module for video recognition,'' in \emph{International Conference on Computer Vision (ICCV)}, 2021, pp. 13\,708--13\,718.

\bibitem{RSTAN}
Q.~Liu, X.~Che, and M.~Bie, ``R-stan: Residual spatial-temporal attention network for action recognition,'' \emph{IEEE Access}, pp. 82\,246--82\,255, 2019.

\bibitem{STA}
Y.~Fu, X.~Wang, Y.~Wei, and T.~Huang, ``Sta: Spatial-temporal attention for large-scale video-based person re-identification,'' in \emph{Proceedings of the AAAI conference on artificial intelligence}, 2019, pp. 8287--8294.

\bibitem{cross-attention}
N.~Carion, F.~Massa, G.~Synnaeve, N.~Usunier, A.~Kirillov, and S.~Zagoruyko, ``End-to-end object detection with transformers,'' in \emph{Computer Vision--ECCV 2020: 16th European Conference, Glasgow, UK, August 23--28, 2020, Proceedings, Part I 16}.\hskip 1em plus 0.5em minus 0.4em\relax Springer, 2020, pp. 213--229.

\bibitem{ReferenceSkerryRyan2018Reference}
R.~Skerry-Ryan, E.~Battenberg, Y.~Xiao, Y.~Wang, D.~Stanton, J.~Shor, R.~Weiss, R.~Clark, and R.~A. Saurous, ``{Towards End-to-end Prosody Transfer for Expressive Speech Synthesis with Tacotron},'' in \emph{International Conference on Machine Learning (ICML)}, 2018, pp. 4693--4702.

\bibitem{cross}
T.~Li, X.~Wang, Q.~Xie, Z.~Wang, M.~Jiang, and L.~Xie, ``{Cross-speaker Emotion Transfer Based On Prosody Compensation for End-to-End Speech Synthesis},'' in \emph{International Speech Communication Association (Interspeech)}, 2022, pp. 5498--5502.

\bibitem{msmvc}
Z.~Wang, X.~Wang, Q.~Xie, T.~Li, L.~Xie, Q.~Tian, and Y.~Wang, ``{MSM-VC:} high-fidelity source style transfer for non-parallel voice conversion by multi-scale style modeling,'' \emph{IEEE/ACM Transactions on Audio, Speech, and Language Processing}, vol.~31, pp. 3883--3895, 2023.

\bibitem{vaswani2017attention}
A.~Vaswani, N.~Shazeer, N.~Parmar, J.~Uszkoreit, L.~Jones, A.~N. Gomez, {\L}.~Kaiser, and I.~Polosukhin, ``Attention is all you need,'' in \emph{Neural Information Processing Systems(NeurIPS)}, 2017.

\bibitem{liurui2021expressive}
R.~Liu, B.~Sisman, G.~Gao, and H.~Li, ``Expressive {TTS} training with frame and style reconstruction loss,'' \emph{IEEE/ACM Transactions on Audio, Speech, and Language Processing}, vol.~29, pp. 1806--1818, 2021.

\bibitem{TacotronWang2017}
Y.~Wang, R.~Skerry-Ryan, D.~Stanton, Y.~Wu, R.~J. Weiss, N.~Jaitly, Z.~Yang, Y.~Xiao, Z.~Chen, S.~Bengio, Q.~Le, Y.~Agiomyrgiannakis, R.~Clark, and R.~A. Saurous, ``{Tacotron: Towards End-to-End Speech Synthesis},'' in \emph{International Speech Communication Association (Interspeech)}, 2017, pp. 4006--4010.

\bibitem{LibriTTS}
H.~Zen, V.~Dang, R.~Clark, Y.~Zhang, R.~J. Weiss, Y.~Jia, Z.~Chen, and Y.~Wu, ``Libritts: A corpus derived from librispeech for text-to-speech,'' in \emph{International Speech Communication Association (Interspeech)}, 2019, pp. 1526--1530.

\bibitem{VCTK}
C.~Veaux, J.~Yamagishi, and K.~MacDonald, ``Cstr vctk corpus: English multi-speaker corpus for cstr voice cloning toolkit.''\hskip 1em plus 0.5em minus 0.4em\relax University of Edinburgh. The Centre for Speech Technology Research (CSTR), 2016.

\bibitem{CMU-Arctic}
J.~Kominek and A.~W. Black, ``The cmu arctic speech databases,'' in \emph{5th ISCA Workshop on Speech Synthesis (SSW 5)}, 2004, pp. 223--224.

\bibitem{hifi}
E.~Bakhturina, V.~Lavrukhin, B.~Ginsburg, and Y.~Zhang, ``Hi-fi multi-speaker english tts dataset,'' \emph{ArXiv}, 2021.

\bibitem{librispeech}
V.~Panayotov, G.~Chen, D.~Povey, and S.~Khudanpur, ``Librispeech: an asr corpus based on public domain audio books,'' in \emph{International conference on acoustics, speech and signal processing (ICASSP)}, 2015, pp. 5206--5210.

\bibitem{voxceleb2}
J.~S. Chung, A.~Nagrani, and A.~Zisserman, ``{VoxCeleb2: Deep Speaker Recognition},'' in \emph{International Speech Communication Association (Interspeech)}, 2018, pp. 1086--1090.

\bibitem{zhou2021seenESD}
K.~Zhou, B.~Sisman, R.~Liu, and H.~Li, ``{Seen and Unseen Emotional Style Transfer for Voice Conversion with a New Emotional Speech Dataset},'' in \emph{International Conference on Acoustics, Speech and Signal Processing (ICASSP)}, 2021, pp. 920--924.

\bibitem{tian2020tfgan}
Q.~Tian, Y.~Chen, Z.~Zhang, H.~Lu, L.~Chen, L.~Xie, and S.~Liu, ``{TFGAN: Time and Frequency Domain based Generative Adversarial Network for High-fidelity Speech Synthesis},'' \emph{Arxiv}, 2020.

\bibitem{s2vc}
J.~hao Lin, Y.~Y. Lin, C.-M. Chien, and H.~yi~Lee, ``S2vc: A framework for any-to-any voice conversion with self-supervised pretrained representations,'' in \emph{International Speech Communication Association (Interspeech)}, 2021, pp. 836--840.

\end{thebibliography}

\vfill

\end{document}